\documentclass[aps,showpacs,floatfix,amsmath,twocolumn]{revtex4}
\usepackage[T1]{fontenc}
\usepackage[latin1]{inputenc}
\usepackage{graphicx}
\usepackage{dcolumn}
\usepackage{bm}

\newcommand{\pep}[1]{\underline{#1}}

\begin{document}

\preprint{GANIL-XXX-TH}

\title{Super-Symmetry transformation for excitation processes}
\author{J\'er\^ome Margueron}
\altaffiliation{Present address: Institut f\"ur Theoretische Physik, Universit\"at 
T\"ubingen, D-72076 T\"ubingen, Germany}
\affiliation{GANIL CEA/DSM - CNRS/IN2P3 BP 5027 F-14076 Caen Cedex 5, France}
\author{Philippe Chomaz}
\affiliation{GANIL CEA/DSM - CNRS/IN2P3 BP 5027 F-14076 Caen Cedex 5, France}

\date{\today}

\begin{abstract}
Quantum Mechanics SUper-SYmmetry (QM-SUSY) provides a general framework for
studies using phenomenological potentials for nucleons (or clusters) 
interacting with a core. 
The SUSY potentials result from the transformation of the mean field potential 
in order to account for the Pauli blocking of the core orbitals. 
In this article, we discuss how other potentials (like external probes or 
residual interactions between the valence nucleons) 
are affected by the SUSY transformation. 
We illustrate how the SUSY transformations induce off-diagonal terms
in coordinate space that play the essential role on the induced transition
probabilities on two examples: the electric operators and Gaussian external 
fields.
We show that excitation operators, doorway states, strength and sum rules are
modified.
\end{abstract}


\maketitle

\section{Introduction}

Almost all branches of many-body physics have developed methods to simplify 
a many-body self-interacting system into a local ``effective'' mean potential 
affecting the pertinent degrees of freedom.
It often provides an adequate starting point for more sophisticated
approaches. For example, phenomenological potentials replacing the
Schr\"{o}dinger equation of N self-interacting particles by a one body
potential whose orbits simulate the experimentally known structure have been
widely used in nuclear physics. Of particular interest are the halo systems
which are often described in terms of valence nucleons interacting with a
core. An elegant way to justify the effective core-nucleon phenomenological
interaction is to invoke a super-symmetric (SUSY) transformation
of the mean-field potential of a N-body 
system~\cite{Wit81,And84,Suk85,Bay87,Coo95}. 
Indeed, since the core is made of nucleons occupying the lowest orbitals of 
the mean-field potential, the halo nucleons cannot fill in
these occupied states because of the Pauli exclusion
principle. SUSY transformations including the forbidden states removal
(States Removal Potential, SRP) and the restoration of phase shifts (Phase
Equivalent Potential, PEP) provide an exact way to remove the states 
occupied by the core without
altering the remaining states properties. Hence, SUSY transformation which
can be fully analytical for some classes of potentials~\cite{Coo95}, provides
an equivalent effective interaction between composite systems and
thus can be safely used to describe nuclear structure and reaction of nuclei
presenting a high degree of clusterization. 

SUSY transformations have been applied to breakup mechanisms involving halo
nuclei~\cite{Rid96,Cap03}. Indeed, the phenomenological treatment of 
halo nuclei in terms of nucleons in interaction with a core
should take into account the fact that some
intrinsic bound states of the nucleon-core
potential are Pauli blocked. For instance, the 
$1s$ orbital is generally occupied by the core nucleons. In the case of a
one neutron halo, like $^{11}$Be or $^{19}$C, SUSY-PEP potentials have been
used to calculate $B$(E1) matrix elements \cite{Rid96}, Coulomb 
breakup~\cite{Cap03}, transfer reactions~\cite{Gon00}. 
In the case of two neutrons halo,
like $^{6}$He, $^{11}$Li, or $^{14}$Be, it has been applied to remove the
forbidden states and to analyze binding energies and radii of these 
nuclei~\cite{Hes99,Tho00,Can01,Des03}. Finally, it has also been included in
coupled-channel calculations \cite{Lee00,Shi00,Sam03}. 
In all these calculations, SUSY transformation has been
applied considering the following approximation: only the internal part of
the Hamiltonian (the core-halo potential) is SUSY transformed while the
additional fields (external potentials, two-body correlations in the halo)
remain unmodified. This approximation will be called the \textit{internal}
SUSY approximation because it concerns only the core-halo potential. 
In the framework of this approximation, the SUSY transformation is not
equivalent to the exact treatment in which the Pauli blocked states are
projected out. In this article, we discuss a consistent SUSY framework which
is always equivalent to the full projection-method.

The accuracy of the \textit{internal} SUSY transformation has been discussed
in several papers. For instance, Ridikas et al.~\cite{Rid96} have analyzed
the radii of several halo nuclei as well as $B$(E1) matrix elements 
before and after the SUSY transformation. Thompson et al.~\cite{Tho00}
and Descouvemont et al.~\cite{Des03} have performed a comparison of the 
full projector-method with the \textit{internal} SUSY. 
As the consistent SUSY framework we discuss is totally equivalent to the full
projection-method, the comparison between the \textit{internal}
approximation and the consistent SUSY treatment is thus an alternative method
to estimate the accuracy of the \textit{internal} approximation. 
From the theoretical point of view, the consistent SUSY approach provides 
a unique and exact framework to compute excitation processes or to 
take into account residual interaction between valence nuclei.
Such a consistent framework is essential to interpret the results
of inverse problems in scattering theory \cite{Cha77}.

This article is organized in the following way. In section \ref{section:1},
we develop a consistent formalism to map the original Hamiltonian into the
SUSY partner Hamiltonian. In the case of a static problem, this mapping is
the usual one, but we will show in section \ref{section:2} that in the case
of an Hamiltonian modified by either an external field or a two-body
interaction (for instance 2 neutrons in the halo), one should transform
these fields into the new space. In section \ref{section:3}, 
we will illustrate the SUSY transformation showing both 
analytical results and numerical implementations for two
potentials which are important in nuclear physics. 
We will then discuss the transformation of external field:
in section \ref{section:4}, the response to an
electric excitation of the general form ${\hat{r}}^{\lambda }\hat{Y}_{LM}$,
and the response to a Gaussian potential in section \ref{section:5}.


\section{SUSY transformation for one body Hamiltonian \label{section:1}}

The application of supersymmetry to Schr\"{o}dinger quantum mechanics 
\cite{Wit81,And84,Suk85,Bay87,Coo95} has shed new light on the problem of
constructing phase-equivalent potentials. In this section, we review the
SUSY transformations which remove a state (SRP) and impose that the phase
shifts are conserved (PEP) \cite{Bay87b,Anc92}. We will introduce mapping
operators which change the Hamiltonian as well as the bound states. Finally
we will present effects on the external potential and residual interaction
of composite systems described through effective Hamiltonians.

\subsection{Initial Hamiltonian: $\hat{h}_0$}

Quantum mechanics SUSY has been extensively studied for one dimensional 
systems. There are
two ways to perform the multi-dimensional generalization depending on the
choice of space coordinates. In three dimensions, one can choose the
Cartesian coordinates, $(x,y,z)$ \cite{And84} or the spherical coordinates, 
$(r,\Omega )$ \cite{Suk85}. 
We choose the latter which is often used for excitation processes.
Hence, the representation of the one particle Hilbert space 
$\mathcal{H}$ is given by a sum over the sub-spaces 
$\mathcal{H}^{l} $ associated to the angular momentum $l$: 
$\mathcal{H}=\mathcal{H}^{0}+\mathcal{H}^{1}+\mathcal{H}^{2}+\dots$

Let us introduce the initial Hamiltonian $\hat{h}_{0}$: 
\begin{equation}
\hat{h}_{0}=\frac{\mathbf{\hat{p}}^{2}}{2m}+\mathrm{\hat{v}}_{0},
\label{EQ:0}
\end{equation}
where $\mathbf{\hat{p}}$ is the momentum operator and where the potential 
operator $\mathrm{\hat{v}}_{0}$ is assumed to be local. 
Since $\hat{h}_{0}$ is
rotationally invariant we can introduce angular momentum as good quantum
number and thus the wave functions associated to an energy $E$ can be
written as $\left| \phi _{0}(E)\right\rangle =\frac{1}{\hat{r}}\left|
\varphi _{0}^{l}(E)\right\rangle \otimes \left| Y_{lm}\right\rangle $. In
the sub-space $\mathcal{H}^{l}$, the radial static Schr\"{o}dinger equation
associated to the $l$th partial wave is
\begin{equation}
\hat{h}_{0}^{l}\left| \varphi _{0}^{l}(E)\right\rangle \equiv \left( 
\frac{\hat{p}^{2}}{2m}+\hat{v}_{0}^{l}\right) \left| \varphi
_{0}^{l}(E)\right\rangle =E\left| \varphi _{0}^{l}(E)\right\rangle,
\end{equation}
where $\hat{p}=i\hbar \;\hat{\nabla}_{r}$ is the radial momentum operator.
The effective radial potential $\hat{v}_{0}^{l}$ includes the centrifugal force 
\begin{equation}
\hat{v}_{0}^{l}=\frac{\hbar ^{2}}{2m}\frac{l(l+1)}{\hat{r}^{2}}+
\mathrm{\hat{v}}_{0}.
\end{equation}
In order to simplify the discussion, we do not include the
spin-orbit potential. Nevertheless, the generalization of this framework to
include spin-orbit potential is not difficult.

In order to simplify the notations when there is no ambiguity we will drop
the label $l$ since the SUSY transformations considered are defined
in a subspace of angular momentum $l$ (and $m)$ $i.e.$ they are
block-diagonal in the complete space. Thus, they affect differently the
potential $\hat{v}^{l}$ associated with different $l$.

\subsection{Hamiltonian after $k$ SUSY transformations: $\hat{h}_k$}

The elementary SUSY transformations remove a single state with or 
without restoring the phase shifts. In order to remove several states 
we will iterate the SUSY transformation.
Therefore, let us assume that after $k$ transformations 
the radial static Schr\"{o}dinger equation associated to the $l$th partial
wave can be written as 
\begin{equation}
\hat{h}_{k}^{l}\left| \varphi _{k}^{l}(E)\right\rangle \equiv \left( 
\frac{\hat{p}^{2}}{2m}+\hat{v}_{k}^{l}\right) \left| \varphi
_{k}^{l}(E)\right\rangle =E\left| \varphi _{k}^{l}(E)\right\rangle.
\label{eq1}
\end{equation}
It should be noticed that, since the SUSY transformations are 
block-diagonal and different in each
subspace of angular momentum $l,$ the different radial potentials do not
correspond to the same potential $i.e.$ the various 
$\mathrm{\hat{v}}_{k}^{l}$ are different 
\begin{equation}
\mathrm{\hat{v}}_{k}^{l}=\hat{v}_{k}^{l}-\frac{\hbar ^{2}}{2m}\frac{l(l+1)}
{\hat{r}^{2}}
\end{equation}
for different angular momenta $l$.
We call $E_{k}^{(i)}$ ($i=n,l$) the energy of the $i$th bound state of 
$\hat{h}_{k}^{l}$ which is thus $(2l+1)$-fold degenerate.

The bound states correspond to the square integrable solutions of the
differential equation (Eq.~\ref{eq1}). However, we will not restrict the
solution of Eq.~\ref{eq1} to bound states but rather consider all solutions 
$\left| \varphi _{k}(E)\right\rangle $. Given a 
particular solution $\left| 
\widetilde{\varphi }_{k}(E)\right\rangle $ of Eq.~\ref{eq1} whose inverse is
square integrable, the general solution of Eq.~\ref{eq1} can be recast 
as~\cite{Suk85} 
\begin{equation}
\varphi _{k}(E,\alpha ;r)=\widetilde{\varphi }_{k}(E;r)\left( 1+\alpha
\int_{r}^{\infty }\frac{dr^{\prime }}{\left( 
\widetilde{\varphi}_{k}(E;r^{\prime })\right)^{2}}\right),  
\label{EQ:sol-general}
\end{equation}
where the parameter $\alpha $ can vary freely to construct all the possible
solutions of the second-order differential equation (Eq.~\ref{eq1}). In 
Eq.~\ref{EQ:sol-general}, we use the r-representation and the Dirac 
notations: 
$\widetilde{\varphi }_{k}(E;r)=\left\langle r\left| \widetilde{\varphi }%
_{k}(E)\right. \right\rangle $. For future use let us define the constant 
$\beta = \Big(\int_{0}^{\infty }dr/[\widetilde{\varphi }_{k}(E;r)]^{2}\Big)^{-1}$.

The Hamiltonians $\hat{h}_{k}$ can always be factorized 
\begin{equation}
\hat{h}_{k}={\hat{a}}_{k}^{+}{\hat{a}}_{k}^{-}+\mathcal{E}_{k}, \label{eq3}
\end{equation}
where $\mathcal{E}_{k}$ is the factorization energy and the first-order
differential operators ${\hat{a}}{_{k}^{±}}$ 
(${\hat{a}}{_{k}^{-}}=\left( {\hat{a}}{_{k}^{+}}\right)^{\dagger}$) 
are of the following form: 
\begin{equation}
{\hat{a}}{_{k}^{\pm}}=\frac{1}{\sqrt{2m}}\left( \hbar \hat{w}_{k}\mp i\hat{p}
\right).  \label{eq4}
\end{equation}
where $\hat{w}_{k}\equiv w_{k}(\hat{r})$ is the super-potential. 
Notice that in the literature, capital letters are usually used for the 
differential operators $\hat{a}_k^\pm$. Here, we dedicate capital letters for
many-body operators while lower case letters are used for one-body
operator. It is possible to show that the general solution 
$\left|\varphi_{k}(\mathcal{E}_{k},\alpha )\right\rangle$ of Eq.~\ref{eq1}
with $E=\mathcal{E}_{k}$ is equivalently the solution of the first order
differential equation 
\begin{equation}
{\hat{a}}_{k}^{-}\left|\varphi_{k}(E=\mathcal{E}_{k},\alpha )\right\rangle=0.  
\label{eq7}
\end{equation}
As a consequence, the super-potential is the local operator defined by 
\begin{equation}
w_{k}(E=\mathcal{E}_{k},\alpha ;r)=\frac{d}{dr}\ln \varphi _{k}
(E=\mathcal{E}_{k},\alpha ;r).  
\label{eq8}
\end{equation}
For a given factorization energy $\mathcal{E}_{k}$, there is a family of
solutions which depends on the parameter $\alpha $ generating the
super-potential $\hat{w}_{k}(E=\mathcal{E}_{k},\alpha )$. Note that 
$\varphi_{k}(E=\mathcal{E}_{k},\alpha ;r)$ must be nodeless in order 
$\hat{a}_{k}^{\pm}$ to be bound. 
Hence $\mathcal{E}_{k}$ must be less than or equal to the
ground state energy $E_{k}$ of $\hat{h}_{k}$ and this requires also that 
$\alpha>-\beta$. The choice of the factorization energy $\mathcal{E}_{k}$
and the selection of a member from the family of solutions $w_{k}$ must
clearly be physically motivated.

In this section we have defined the notations used in the following. In the
next section we will present a 2 step method which removes the lowest-energy
state and preserves the phase shifts.

\subsection{State Removal Potential (SRP): $\hat{v}_{k+1}$}

The SRP transformation is defined so that it removes the lowest-energy state
of a given sub-space $\mathcal{H}^{l}$. For the given angular momentum $l$,
we choose $\mathcal{E}_{k}=E_{k}^{0}$, the energy of the lowest energy
state of the
Hamiltonian $\hat{h}_{k}$. It follows that the inverse of the particular
solution is not square integrable and it imposes $\alpha$=0. 
With these definitions, we associate to $\hat{h}_{k}$ a supersymmetric
partner $\hat{h}_{k+1}$ defined by 
\begin{eqnarray}
\hat{h}_{k+1} &=&{\hat{a}}_{k}^{-}{\hat{a}}_{k}^{+}+\mathcal{E}_{k}
=\frac{\hat{p}^{2}}{2m}+\hat{v}_{k+1}, \\
\hat{v}_{k+1} &=&\hat{v}_{k}-\frac{\hbar ^{2}}{m}\left( \partial _{r}\;
\hat{w}_{k}(\mathcal{E}_{k},\alpha =0)\right).  \label{eq11}
\end{eqnarray}
The Hamiltonians $\hat{h}_{k}$ and $\hat{h}_{k+1}$ share the same spectrum
except for the lowest-energy state of $\hat{h}_{k}$ which has been suppressed in 
$\hat{h}_{k+1}$. The states ($\left| \varphi _{k+1}(E)\right\rangle $) of 
$\hat{h}_{k+1}$ can be obtained from those 
($\left|\varphi_{k}(E)\right\rangle $) of $\hat{h}_{k}$ according to 
\begin{equation}
\left| \varphi _{k+1}(E)\right\rangle =\frac{1}{\sqrt{\hat{h}_{k+1}-
\mathcal{E}_{k}}}{\hat{a}}{_{k}^{-}}\left| \varphi _{k}(E)\right\rangle \equiv 
{\mathrm{\hat{u}}}_{k}^{-}\left| \varphi _{k}(E)\right\rangle.
\end{equation}
Conversely, except for the ground state, the states of $\hat{h}_{k}$ can be 
obtained from those of $\hat{h}_{k+1}$ by 
\[
\left| \varphi _{k}(E)\right\rangle =\frac{1}{\sqrt{\hat{h}_{k}-
\mathcal{E}_{k}}}{\hat{a}}{_{k}^{+}}\left| \varphi _{k+1}(E)\right\rangle 
={ \mathrm{\hat{u}}}_{k}^{+}\left| \varphi_{k+1}(E)\right\rangle .
\]
In these equations, we have introduced the pseudo unitary SRP-operators 
$\hat{\mathrm{u}}{_{k}^{-}}$ and $\hat{\mathrm{u}}{_{k}^{+}}$ which are
defined as (the products ${\hat{a}}{_{k}^{+}}{\hat{a}}{_{k}^{-}}$ and 
${\hat{a}}{_{k}^{-}}{\hat{a}}{_{k}^{+}}$ being definite positive) 
\begin{eqnarray}
\mathrm{\ }\hat{\mathrm{u}}{_{k}^{+}} &=&{\hat{a}}{_{k}^{+}}\frac{1}
{\sqrt{{\hat{a}}{_{k}^{-}}{\hat{a}}{_{k}^{+}}}}=\frac{1}{\sqrt{{\hat{a}}{_{k}^{+}}
{\hat{a}}{_{k}^{-}}}}{\hat{a}}{_{k}^{+}},  \label{eq34} \\
\mathrm{\ }\hat{\mathrm{u}}{_{k}^{-}} &=&{\hat{a}}{_{k}^{-}}\frac{1}
{\sqrt{{\hat{a}}{_{k}^{+}}{\hat{a}}{_{k}^{-}}}}=\frac{1}
{\sqrt{{\hat{a}}{_{k}^{-}}{\hat{a}}{_{k}^{+}}}}{\hat{a}}{_{k}^{-}}.  
\label{eq35}
\end{eqnarray}
These operators are pseudo-unitary since 
$\hat{\mathrm{u}}{_{k}^{-}=(\hat{\mathrm{u}}{_{k}^{+}})}^{\dagger }$, 
$\hat{\mathrm{u}}{_{k}^{-}}\hat{\mathrm{u}}{_{k}^{+}}=\hat{1}$ 
and $\hat{\mathrm{u}}{_{k}^{+}}\hat{\mathrm{u}}{_{k}^{-}}=\hat{p}$, 
where the projector $\hat{p}$ suppresses the
lowest-energy state $|\varphi _{k}^{0}\rangle $ of the Hamiltonian 
$\hat{h}_{k}$ from the sub-space $\mathcal{H}^{l}$ and can be written as 
$\hat{p}=1-|\varphi_{k}^{0}\rangle \langle \varphi _{k}^{0}|$.

The relation between $\hat{h}_{k}$ and $\hat{h}_{k+1}$ is 
\begin{equation}
\hat{h}_{k+1}=\frac{\hat{p}^{2}}{2m}+\hat{v}_{k+1}=\hat{\mathrm{u}}_{k}^{+}
\,\left( \frac{\hat{p}^{2}}{2m}+\hat{v}_{k}\right) \,\hat{\mathrm{u}}_{k}^{-}
=\hat{\mathrm{u}}_{k}^{+}\,\hat{h}_{k}\,\hat{\mathrm{u}}_{k}^{-}
\label{eq35b}
\end{equation}
However, it is important to remark that 
$\hat{v}_{k+1}\neq \hat{\mathrm{u}}_{k}^{+}\hat{v}_{k}\hat{\mathrm{u}}_{k}^{-}$ 
and $\hat{p}^{2}\neq \hat{\mathrm{u}}_{k}^{+}\hat{p}^{2}\hat{\mathrm{u}}_{k}^{-}$. 
In fact the SUSY
transformation of a local potential is not local. The simple
diagonal form of the potential Eq.~\ref{eq11} is recovered because, by
construction, the modifications of the kinetic part just cancel the
off-diagonal terms in the transformed potential. Then, the kinetic and the
potential parts of the Hamiltonian should be transformed together in order to
get the relation (\ref{eq35b}) with a simple potential (local in the 
$r$-space) and kinetic (diagonal in the $p$-representation) terms.

\subsection{Phase Equivalent Potential (PEP): ${\pep{v}}_{k+1}$}

It can be shown that SRP transformations change the phase shifts. To solve
this problem, Baye \cite{Bay87} has proposed to perform a second SUSY
transformation and associate to $\hat{h}_{k+1}$ a new supersymmetric partner 
$\hat{{\pep{h}}}_{k+1}$ so that 
\begin{equation}
\hat{{\pep{h}}}_{k+1}=\hat{{\pep{a}}}_{k}^{-}\hat{\pep{a}}_{k}^{+}+
{\pep{{\cal E}}}_{k}=\frac{\hat{p}^{2}}{2m}+\hat{\pep{v}}_{k+1}, \\
,
\end{equation}
with ${\pep{{\cal E}}}_{k}=E_{k}^{0}$, the ground state energy of $\hat{h}_k$,
and $\alpha =-\beta $. 
In this case, the solution $\pep{\varphi}_{k+1}$ of $\hat{h}_{k+1}$
and its inverse are not square integrable. 
The second SUSY transformation does not suppress nor add any state to the
spectrum of $\hat{h}_{k+1}$, but it restores the phase shifts so that the
Hamiltonian $\hat{{\pep{h}}}_{k+1}$ is equivalent to $\hat{h}_{k}$ as far as the
scattering properties are concerned. Note that the energy ${\pep{{\cal E}}}_{k}=E_{k}^{0}$ 
is now below the ground state energy $E_{k+1}^{0}$ of 
$\hat{h}_{k+1}$.

The corresponding super-potential $\hat{{\pep{w}}}_{k}({\pep{{\cal E}}}_{k})$ is
deduced from the ground state wave function $|\varphi_k^0\rangle$ of 
$\hat{h}_{k+1}$ according to Eq.~\ref{eq8}. 
It can also be deduced from the ground state of $\hat{h}_{k}$ according to \cite{Bay87b}: 
\begin{eqnarray}
{\pep{w}}_{k}({\pep{{\cal E}}}_{k};r)&=&\frac{d}{dr}\ln \frac{1}{\varphi_k^0(r)}
\int_0^r dr^\prime \, \left( \varphi_k^0(r^\prime)\right)^2 
\nonumber \\
&=&\underline{\pep{w}}_{k}({\pep{{\cal E}}}_{k};r)-w_k({\mathcal{E}}_{k};r),
\end{eqnarray}
where we have used the relation $\pep{\cal E}_k=\mathcal{E}_k$ and introduced
the modified-super-potential $\underline{{\pep{w}}}_{k}({\pep{{\cal E}}}_{k};r)$ as 
\begin{eqnarray}
\underline{\pep{w}}_{k}({\pep{{\cal E}}}_{k};r)=\frac{d}{dr}\ln \int_0^r dr^\prime
\, \left( \varphi_k^0(r^\prime)\right)^2.
\end{eqnarray}

The corresponding potential $\hat{{\pep{v}}}_{k+1}$ is 
\begin{eqnarray}
\hat{{\pep{v}}}_{k+1}&=&\hat{v}_{k+1}-\frac{\hbar^2}{m}\left(\partial_r\;\hat{\pep{w}}_{k}
({\pep{{\cal E}}}_{k})\right) \nonumber \\
&=&\hat{v}_{k}-\frac{\hbar^2}{m}\left(\partial_r\;\hat{\underline{\pep{w}}}_{k}
({\pep{{\cal E}}}_{k})\right).
\end{eqnarray}

According to the above discussion, the spectra of $\hat{h}_{k+1}$ and 
$\hat{{\pep{h}}}_{k+1}$ are identical. All the states of $\hat{h}_{k}$, except its
lowest-energy state, are mapped onto the states of $\hat{{\pep{h}}}_{k+1}$ with the
same phase shifts. This mapping is simply: 
\begin{eqnarray}
\left| \pep{\varphi}_{k+1}(E)\right\rangle \!\!&=&\!\!\frac{1}{E_{k}^{0}-\hat{{\pep{h}}}_{k+1}}
{\hat{\pep{a}}}{_{k}^{-}}{\hat{a}}{_{k}^{-}}\left| \varphi _{k}(E)\right\rangle
\equiv \hat{\mathrm{\pep{u}}}{_{k}^{-}}\left| \varphi _{k}^{l}(E)\right\rangle,
\nonumber \\
\label{eq24} \\
\left| \varphi_{k}(E)\right\rangle \!\!&=& \!\!\frac{1}{E_{k}^{0}-
\hat{h}_{k}}\hat{a}{_{k}^{+}}{\hat{\pep{a}}}{_{k}^{+}}\left| 
\pep{\varphi}_{k+1}(E)\right\rangle =\hat{\mathrm{\pep{u}}}{_{k}^{+}}\left| 
\pep{\varphi}_{k+1}(E)\right\rangle,
\nonumber \\
\end{eqnarray}
with the pseudo-unitary PEP-operators 
\begin{eqnarray}
\hat{\mathrm{\pep{u}}}_{k}^{+} &=&-\hat{a}_{k}^{+}\hat{\pep{a}}_{k}^{+}
\frac{1}{\hat{\pep{a}}_{k}^{-}\hat{\pep{a}}_{k}^{+}}=-\frac{1}{\hat{a}_{k}^{+}
\hat{a}_{k}^{-}}\hat{a}_{k}^{+}\hat{\pep{a}}_{k}^{+},
\nonumber \\
\label{eq40} \\
\hat{\mathrm{\pep{u}}}_{k}^{-} &=&-\frac{1}{\hat{\pep{a}}_{k}^{-}\hat{\pep{a}}_{k}^{+}}
\hat{\pep{a}}_{k}^{-}\hat{a}_{k}^{-}=-\hat{\pep{a}}_{k}^{-}\hat{a}_{k}^{-}
\frac{1}{\hat{a}_{k}^{-}\hat{a}_{k}^{+}}, \nonumber \\
\label{eq41}
\end{eqnarray}

The relation between $\hat{h}_{k}$ and $\hat{\pep{h}}_{k+1}$ is 
\begin{equation}
\hat{\pep{h}}_{k+1}=\hat{\mathrm{\pep{u}}}_{k}^{+}\,\hat{h}_{k}\,\,
\hat{\mathrm{\pep{u}}}_{k}^{-}. \label{EQ:h2}
\end{equation}

The advantage of using the operators $\hat{u}_k^\pm$ and $\hat{\pep{u}}_k^\pm$
is that all the relations we will deduce hereafter will be algebraically
equivalent for SRP and PEP transformations.
In the following, as long as no confusion is possible, we will write 
the relations fulfilled by the general operator $\hat{u}_k^\pm$,
which can be replaced either by the operator $\hat{u}_k^\pm$ for the
SRP transformation or $\hat{\pep{u}}_k^\pm$ for the PEP one.

 \section{SUSY transformation for general Hamiltonians \label{section:2}}

In nuclear physics, we are often interested in the description of 
$\mathrm{A}$ interacting nucleons assuming that these nucleons can be
separated into a frozen core containing $\mathrm{A}_{\mathrm{c}}$ nucleons
and a valence space containing $\mathrm{A}_{\mathrm{v}}$ nucleons. Hence,
the wave function of this system is assumed to factorize into a core and a
valence part, $\left| \Phi (\mathrm{A}_{\mathrm{c}}+\mathrm{A}_{\mathrm{v}%
})\right\rangle =\left| \Phi _{\mathrm{c}}(\mathrm{A}_{\mathrm{c}%
})\right\rangle \otimes \left| \Phi _{\mathrm{v}}(\mathrm{A}_{\mathrm{v}%
})\right\rangle $. The core state is described at the mean field 
level as a Slater determinant
of $\mathrm{A}_{\mathrm{c}}$ single particle states 
$\left| \phi _{\mathrm{c}}(n)\right\rangle$ 
occupying the $\mathrm{A}_{\mathrm{c}}$ 
lowest-energy eigenstates of the mean field potential $\hat{h}_{0}$: 
$\left| \Phi _{\mathrm{c}}(\mathrm{A}_{\mathrm{c}})\right\rangle =\widetilde{%
\prod_{n=1}^{\mathrm{A}_{\mathrm{c}}}\left| \phi _{c}(n)\right\rangle }$
where $\widetilde{...}$ stands for the antisymmetrization sign. As a
consequence of the Pauli principle, the valence nucleons cannot occupy the
lowest orbitals of the core-valence potential which are already occupied by
the core nucleons.  The evolution of 
$\left| \Phi _{\mathrm{v}}(\mathrm{A}_{\mathrm{v}})\right\rangle$ 
is thus ruled by the Hamiltonian $\hat{H}_{\mathrm{v}}$ 
which contains a projection out of the occupied space 
$\hat{\mathrm{P}}_{\mathrm{v}}=\prod_{n=1}^{\mathrm{A}_{\mathrm{c}}}%
\hat{c}_{n}^{{}}$ $\hat{c}_{n}^{+}$ where 
$\hat{c}_{n}^{+}$ ($\hat{c}_{n}^{{}}$) is the creation (annihilation)
operator of a nucleon in the occupied orbital 
$\left| \phi _{\mathrm{c}}(n)\right\rangle$. 
$\hat{H}_{\mathrm{v}}$ is assumed to contain
the confining effect of the mean field $\hat{h}_{0}$. For cases with several
nucleons in the valence space, the residual interaction among valence
nucleons, $\hat{V}_{0}$ should
be taken into account when the problem of correlations is addressed. 
Finally, an external field, $\hat{f}_{0}$, should be introduced in order 
to compute excitation properties. Then the Hamiltonian reads 
\begin{equation}
\hat{H}_{\mathrm{v}}=\hat{\mathrm{P}}_{\mathrm{v}}
\hat{H}\hat{\mathrm{P}}_{\mathrm{v}}
\end{equation}
with
\begin{equation}
\hat{H}=\sum_{i=1}^{\mathrm{A}_{\mathrm{v}}}\hat{h}_{0}(i)
+\frac{1}{2}\sum_{i,j=1}^{\mathrm{A}_{\mathrm{v}}}\hat{V}_{0}(i,j)
+\sum_{i=1}^{\mathrm{A}_{\mathrm{v}}}\hat{f}_{0}(i). 
\label{val01}
\end{equation}

In the following, we propose to generalize the SUSY transformation so that
it remains totally equivalent to the projector method for every kind of
additional potentials. The first step of this method is to remove the $k$
orbitals occupied by the core nucleons. 
We introduce the full operator $\hat{\mathrm{U}}_{k-1}^\pm$ which is the
product of the different transformations $\hat{\mathrm{u}}_{k-1}^\pm$
(c.f. Eq.~\ref{eq34} and Eq.~\ref{eq40})
\begin{equation}
\hat{\mathrm{U}}_{k-1}^{^\pm}=\prod_{i=1}^{\mathrm{A}_{\mathrm{v}}}
\hat{\mathrm{u}}_{k-1}^{^\pm}(i)
\label{fullop}
\end{equation}
applying to each single particle wave function $(i)$ of the valence state.
Since those different transformations affects only a given single-particle 
angular-momentum subspace, the total operator 
$\hat{\mathrm{U}}_{k}^{^{\pm}}$ is block diagonal in spin representation. 
Being the product of pseudo-unitary transformations, 
$\hat{\mathrm{U}}_{k}^{^{\pm}}$ is also pseudo-unitary since 
$\hat{\mathrm{U}}_{k}^{-}=\left( \hat{\mathrm{U}}_{k}^{+}\right) ^{\dagger }$, 
$\hat{\mathrm{U}}_{k}^{-}\hat{\mathrm{U}}_{k}^{+}=\hat{1}$ and 
$\hat{\mathrm{U}}_{k}^{+}\hat{\mathrm{U}}_{k}^{-}=\hat{\mathrm{P}}_{\mathrm{v}}$.
Using $\hat{\mathrm{U}}_{k}^{+}\hat{\mathrm{U}}_{k}^{-}=\mathrm{\hat{P}}_{\mathrm{v}}$, 
we can thus write 
$\hat{H}_{\mathrm{v}}=\mathrm{\hat{P}}_{\mathrm{v}}\hat{H}\mathrm{\hat{P}}_{\mathrm{v}}$
and explicitly
\begin{equation}
\hat{H}_{\mathrm{v}}=\hat{\mathrm{U}}_{k-1}^{+}
\hat{\mathrm{U}}_{k-1}^{-}H\hat{\mathrm{U}}_{k-1}^{+}\hat{\mathrm{U}}_{k-1}^{-}=
\hat{\mathrm{U}}_{k-1}^{+}\hat{H}_{\mathrm{v},k}\hat{\mathrm{U}}_{k-1}^{-} 
\end{equation}
where we have introduced the transformed Hamiltonian
\begin{eqnarray}
\hat{H}_{\mathrm{v},k}&=&\hat{\mathrm{U}}_{k-1}^{-}
\hat{H}\hat{\mathrm{U}}_{k-1}^{+} \nonumber \\
&=&\sum_{i=1}^{\mathrm{A}_{\mathrm{v}}}\hat{h}_{k}(i)+
\frac{1}{2}\sum_{i,j=1}^{\mathrm{A}_{\mathrm{v}}}\hat{V}_{k}(i,j)+
\sum_{i=1}^{\mathrm{A}_{\mathrm{v}}}\hat{f}_{k}(i). 
\label{EQ:SUSY-H}
\end{eqnarray}
It is clear from this relation that not only $\hat{h}_{0}$ is
transformed but also the two body interaction is changed into 
$\hat{V}_{k}(i,j)$. Using Eq.~\ref{fullop} and 
$\hat{\mathrm{u}}_{k}^{-}\left( i\right) \hat{\mathrm{u}}_{k}^{+}\left( i\right)%
=\hat{1}\left( i\right)$ we get 
\begin{eqnarray}
\hat{V}_{k}(i,j) &=&\hat{\mathrm{U}}_{k-1}^{-}\hat{V}_{0}(i,j)
\hat{\mathrm{U}}_{k-1}^{+} \label{tbi} \\
&=&\hat{\mathrm{u}}_{k-1}^{-}(i)\hat{\mathrm{u}}_{k-1}^{-}(j)\hat{V}_{0}(i,j)
\hat{\mathrm{u}}_{k-1}^{+}(i)\hat{\mathrm{u}}_{k-1}^{+}(j)
\nonumber \\
\end{eqnarray}
and the external potential $\hat{f}_{0}$ is mapped into 
\begin{equation}
\hat{f}_{k}(i)=\hat{\mathrm{U}}_{k-1}^{-}\hat{f}_{0}(i)
\hat{\mathrm{U}}_{k-1}^{+}=
\hat{\mathrm{u}}_{k-1}^{-}(i)\hat{f}_{0}(i)\hat{\mathrm{u}}_{k-1}^{+}(i).  
\label{obi}
\end{equation}
Note that when $\hat{f}_{0}$ is not a scalar operator, the mapping operator 
$\hat{\mathrm{u}}_{k-1}^\pm$ on the right and on the left of Eq.~\ref{obi} 
may not correspond to the same angular momentum $l$.

It should also be noticed that not only the Hamiltonian is changed but also
the wave functions since the state of the valence nucleons is transformed
into 
\begin{equation}
\left| \Phi _{\mathrm{v},k}\left( t\right) \right\rangle =
\hat{\mathrm{U}}_{k-1}^{-}\left| \Phi _{\mathrm{v}}\left( t\right) \right\rangle.
\label{EQ:SUSY-WF}
\end{equation}

The evolution of a state $|\Phi _{\mathrm{v}}(t)\rangle $ is driven by the
time dependent Schr\"{o}dinger equation 
\[
i\hbar \frac{d}{dt}\left| \Phi _{\mathrm{v}}(t)\right\rangle =
\hat{H}_{\mathrm{v}}\left| \Phi _{\mathrm{v}}(t)\right\rangle ,
\]
which can be mapped into the new Hilbert space where the Pauli-blocked
states have been removed within SUSY transformations: 
\[
i\hbar \frac{d}{dt}\left| \Phi _{\mathrm{v},k}(t)\right\rangle =
\hat{H}_{\mathrm{v},k}\left| \Phi _{\mathrm{v},k}(t)\right\rangle .
\]
It is important to remark that the projectors $\hat{P}_{\mathrm{v}}$ involved in
the definition of the valence Hamiltonian (cf Eq.~\ref{val01}) have been
removed in the mapped Hamiltonian $\hat{H}_{\mathrm{v},k}$ 
(cf Eq.~\ref{EQ:SUSY-H}). 
Hence, the time dependent Schr\"{o}dinger equation in the SUSY
space is simpler than the original
Schr\"{o}dinger equation which involves projection operators.
Nevertheless, the two Schr\"{o}dinger equations written in the original
space or in the SUSY transformed Hilbert space contain strictly the
same physical ingredients and are mathematically equivalent.

In the literature, the transformation of both the excitation operators 
and the wave functions are
usually neglected i.e. $\hat{f}_{0}$ is
often used instead of $\hat{f}_{k}$ and the wave functions are not 
transformed back when evaluating 
observables~\cite{Rid96,Cap03,Gon00,Hes99,Tho00,Can01,Lee00,Shi00,Sam03}. 
In the following, we will study a particularly important application which
is the evaluation of the response of the nucleus to an external (one body)
perturbation (time dependent or not). The use of a SUSY 
transformed two body residual-interaction in the calculation of 
correlations and reactions will be the subject of forthcoming 
studies.

\section{Examples of SUSY transformations \label{section:3}}

In this section, we illustrate the formalism developed above
with two important physical examples: i) the harmonic oscillator potential
which is mostly analytical, it allows a deeper insight into the 
formalism and provides numerical tests, and
ii) the halo nuclei potential which are of important physical 
interest but can be treated only numerically 
since only asymptotic relations can be deduced analytically.

\subsection{The harmonic oscillator potential \label{pot:ho}}

The harmonic oscillator potential is a textbook example \cite{Law80}.
We set
the local potential to be $V_{0}(r)=-V_{0}+\frac{\hbar ^{2}}{2m}(r/b)^{2}$
with $b^{2}=\hbar /m\omega $. In the following, we introduce a reduced
coordinate $q=r/b.$

The eigenstates are labeled with the quantum numbers $(n,l)$ and are
associated with a set of energies $E_{nl}=-V_{0}+(2n+l+3/2)\hbar \omega $. For
each $l$ the lowest energy state is
\[
\varphi _{0}^{0l}(q)=c_{l}\,q^{l+1}\,\exp \{-\frac{1}{2}q^{2}\} ,
\]
with $c_{l}=b^{l-1/2}/\pi ^{3/4}$. 

We deduce the super-potentials for SRP and PEP transformations: 
\begin{eqnarray}
w_{0}^{l}(q) &=&\frac{1}{b}\left( \frac{l+1}{q}-q\right), \\
\underline{\pep{w}}_{0}^{l}(q) &=&\frac{1}{b}\frac{q^{2l+2}\,{\hbox{e}}^{-q^{2}}}
{\mathrm{erf}(q,2l+2)},
\end{eqnarray}
with the unnormalized error function defined by 
$\mathrm{erf(z,l)=\int_{0}^{z}t^{l}\hbox{e}^{-t^{2}}dt}$. 
The differential operators $\hat{a}_{k}^{l^{\pm}}$ are 
\begin{eqnarray}
\langle r|\hat{a}_{0}^{l^{\pm}}|r'\rangle 
&=&-\langle r|\frac{1}{\sqrt{2}}\left( \hat{\mathrm{r}} \pm i
\hat{\mathrm{p}}-\hbar \omega \frac{l+1}{\hat{\mathrm{r}}}\right)|r'\rangle, \\
\langle r|\hat{\pep{a}}_{0}^{l^{\pm}} |r^{\prime }\rangle &\stackrel{\infty }{\sim }
&-\langle r|\frac{1}{\sqrt{2}}\left( -\hat{\mathrm{r}}\pm i\hat{\mathrm{p}}
\right) |r^{\prime }\rangle,
\end{eqnarray}
where $\hat{\mathrm{r}}=\sqrt{m}\omega \hat{r}$ and 
$\hat{\mathrm{p}}=\hat{p}/\sqrt{m}$.
Notice that PEP transformation is not physical in this case because there are
no phase for the harmonic oscillator potential. 
Nevertheless, it remains interesting for the discussion.

From the expressions of the super-potentials removing only one state,
we deduce the transformed potentials: 
\begin{eqnarray}
v_{1}^{l}(r) &=&v_{0}^{l}(r)+\hbar \omega +\frac{\hbar ^{2}(l+1)}{mr^{2}}
=v_{0}^{l+1}(r)+\hbar \omega,  \label{eq:34} \\
{\pep{v}}_{1}^{l}(r) &=&v_{0}^{l}(r)+\frac{\hbar ^{2}}{mb^{2}}\frac{q^{2l+1}
{\hbox{e}}^{-q^{2}}}{\left( {\mathrm{erf}}(q,2l+2)\right) ^{2}} \times 
\nonumber \\
&&\left( (2l+2-2q^{2}){\mathrm{erf}}(q,2l+2)+q^{2l+3}\right).  \label{eq:35}
\end{eqnarray}

\begin{figure}[htb]
\center
\includegraphics[scale=0.3]{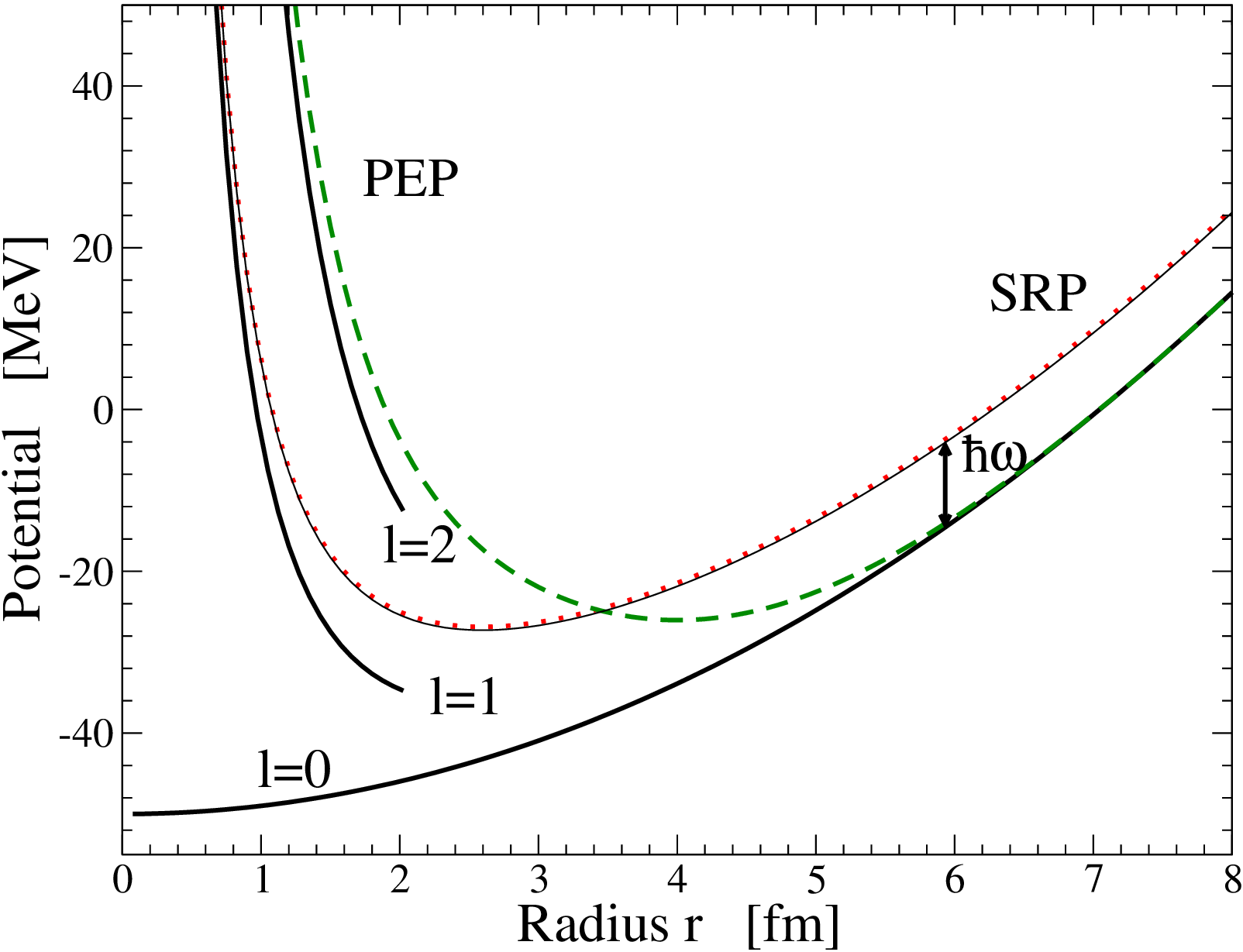}
\caption{(color online) Radial part of the Harmonic Oscillator potential (H-O) for l=0,1,2
(thick lines) compared with its SRP (dotted line) and PEP (dashed
line) transformation of the l=0 potential. The thin solid line stands for 
$v_{\mathrm{H-O}}(l=1)+\hbar \omega $.}
\label{figu01}
\end{figure}

These potentials are represented in Fig.~\ref{figu01}. In the graphical
illustrations we will use nuclear physics scales by taking the following
parameters: $V_{0}$=50 MeV and $\hbar \omega $=10 MeV. The lowest energy
state is at -35 MeV. The r.h.s. of Eq.~\ref{eq:34} demonstrates that the SRP
transformations removing only one state 
have mapped the original potential $v_{0}^{l}$ into a new
potential which is simply $v_{0}^{l^{\prime }}+\hbar \omega $ where the
effective angular momentum is $l^{\prime }=l+1$. 
This is illustrated in Fig.~\ref{figu01} where we have represented the
original potential with $l=0$, $l=1$ and $l=2$ (thick lines) and the
SRP potential obtained numerically (dotted line). These numerical results
have been obtained on a mesh containing 400 points, ranging from 0 fm to 20
fm and with a vanishing boundary condition. The thin solid line is the
analytical result given by Eq.~\ref{eq:34}. The slight difference between
the thin solid line and the dotted line gives an estimate of 
the error of the numerical algorithm which appears to be very small. 

Generalizing this result to the removal of several core states
we remark that, in the new radial Hilbert space, up to a
translation of $n_{c}\,\hbar \omega ,$ where $n_{c}$ is the number of
removed core orbitals with the angular momentum $l,$ the Pauli principle
maps the original potential with angular momentum $l$ to a new potential
analogous to the radial potential with an effective angular momentum 
$l^{\prime }=l+n_{c}$. However, only the radial wave function is affected by
the effective angular momentum, the angular part of the wave function is
unchanged by the SUSY mapping.

As we have already mentioned, this SRP transformation changes the 
phases. 
The restoration of the phases is ensured by the PEP transformation. From the
analytic expression of ${\pep{v}}_{1}^{l}$ (Eq.~\ref{eq:35}), we see that near 
$r\sim 0$ , ${\pep{v}}_{1}^{l}(r)\stackrel{0}{\sim }v_{0}^{l+2}(r)$, and
asymptotically, ${\pep{v}}_{1}^{l}(r)\stackrel{\infty }{\sim }v_{0}^{l}(r)$. The
potential ${\pep{v}}_{1}^{l}(r)$ is represented in Fig.~\ref{figu01} (dashed line).
The restoration of the phases imposes a non trivial transformation of the
potential: near zero, the potential ${\pep{v}}_{1}$ is mapped to a new potential 
analogous to 
$v_{0}^{l^{\prime }}$ with 
a centrifugal force analogous to an effective angular momentum 
$l^{\prime}=l+2n_{c}$, and asymptotically, 
the potential stays unchanged as required by the phase conservation. 
This behavior is the consequence of the Pauli principle and phase
restoration.

The mapping operators, $\hat{\mathrm{u}}_{0}^{l\,±}$,
can also be analytically derived, and we will discuss the properties of
these operators from their asymptotic (all radii going to infinity) 
expressions: 
\begin{eqnarray}
\langle r|\mathrm{u}_{0}^{l+}|r^{\prime }\rangle &\stackrel{\infty }{\sim }
&-\langle r|\frac{1/\sqrt{2}}{\sqrt{\hat{p}^2/2m-E_{0}^{0}}}\left( 
i\hat{\mathrm{p}}+\hat{\mathrm{r}}\right) |r^{\prime }\rangle \\
\langle r|\mathrm{\pep{u}}_{0}^{l\,\pm}|r^{\prime }\rangle &\stackrel{\infty }{\sim}
&\delta (r-r^{\prime })
\end{eqnarray}
Hence, while the operator $\hat{\mathrm{u}}_{0}^{l\,\pm}$ is never trivial
even at large distances the operator $\hat{\mathrm{\pep{u}}}_{0}^{l\,\pm}$ is simply
the unity operator for large r. This is a consequence of the phase
restoration. 
As a consequence, the PEP transformations do not modify observables which
are only sensitive to the asymptotic part of the wave functions. These
asymptotic properties are also valid for other potentials as illustrated for
halo nuclei potential in the next paragraph.

\subsection{Halo nuclei potentials \label{pot:halo}}

The study of the properties of weakly bound systems has found a renewed
interest after the discovery of halo nuclei~\cite{tan85}. These systems have
very large mean square radii and small separation energies. In fact, the
separation energy of the nucleons forming the halo is so small that their
degrees of freedom can be separated from those of the nucleons constituents
of the core. Up to now, this
property has been observed in light nuclei close to the nucleon 
drip-lines like $^{6}$He, $^{11}$Be or $^{19}$C. 
In reference \cite{Gre97}, the proposed core-halo potential for $^{11}$Be is
the sum of a Wood-Saxon potential and a surface potential 
\[
v_{0}^{l=0}(r)=v_{0}f(r)+16a^{2}\alpha \left( \frac{df(r)}{dr}\right) ^{2}, 
\]
where $f(r)$ is a Wood-Saxon potential and the parameters are: $v_{0}=-44.1$
MeV, $\alpha =-10.15$ MeV, $r_{0}^{1}=1.27$ fm and $a_{0}=0.75$ fm. 
The bound states of this potential are $2$ $s$-states at -25.0 MeV and -0.5 MeV
and 1 p-state at -11 MeV. 
For simplicity we omit the
spin-orbit coupling and consider a model case where the 1s and 1p orbitals 
are occupied by the core
neutrons. Thus, these two orbitals are Pauli blocked and cannot be filled in 
by the neutron of the halo. In its ground state the latter occupies 
the $2s$ state.

\begin{figure}[htb]
\center
\includegraphics[scale=0.3]{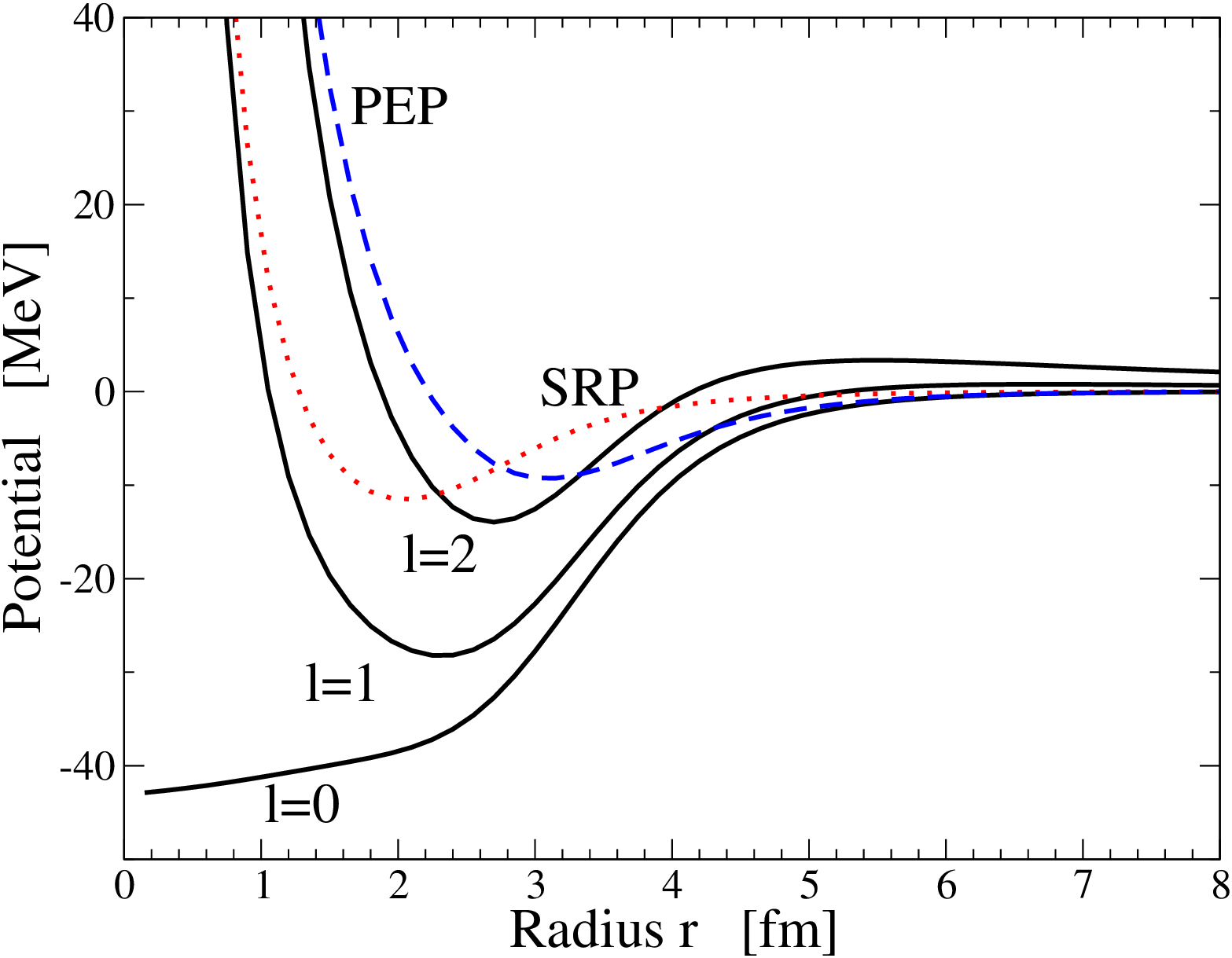}
\caption{(color online) Radial part of the core-halo potentials $v_{0}^{l}$ (solid 
lines)
for $l=0,$ $1$ and $2$ and the SUSY transformed $v_{1}^{l=0}$ (dashed line)
and ${\pep{v}}_{1}^{l=0}$ (dotted line). }
\label{figu02}
\end{figure}

We work on a constant step mesh containing 400 points and ranging from 0 fm
to 50 fm. We show in Fig.~\ref{figu12} the original potential 
$v_{0}^{l} $ for $l=0$, $l=1$ and $l=2$ (solid lines), the SRP 
$v_{1}^{l=0}$ (dashed line) and the PEP ${\pep{v}}_{1}^{l=0}$ (dotted line).

We can obtain analytical expressions near $r=0$ and for large $r$. 
Indeed, near zero, the lowest energy state behaves like $r^{l+1}$ and
asymptotically, it behaves like $\exp (-\gamma _{0}r)$ with 
$\gamma _{0}=\sqrt{-2mE_{0}^{0l}}/\hbar $. 
These asymptotics and therefore the
following expressions are very general
for all potentials which are regular at the origin and 
vanish for large $r$. 
We find that the super-potentials behave like ($r\rightarrow 0$)
\begin{eqnarray}
w_{0}^{l}(r) &\stackrel{0}{\sim }&\frac{l+1}{r}\hbox{ and }w_{0}^{l}(r)
\stackrel{\infty }{\sim }-\gamma _{0}, \\
\underline{\pep{w}}_{0}^{l}(r) &\stackrel{0}{\sim }&\frac{2l+3}{r}\hbox{ and }
\underline{\pep{w}}_{0}^{l}(r)\stackrel{\infty }{\sim }0,
\end{eqnarray}
and the potentials are for $r\rightarrow 0$
\begin{eqnarray}
v_{1}^{l}(r) &\stackrel{0}{\sim }&v_{0}^{l}(r)+\frac{\hbar ^{2}}{m}
\frac{l+1}{r^{2}}=v_{0}^{l+1}(r), \\
{\pep{v}}_{1}^{l}(r) &\stackrel{0}{\sim }&v_{0}^{l}(r)+\frac{\hbar ^{2}}{m}
\frac{2l+3}{r^{2}}=v_{0}^{l+2}(r), 
\end{eqnarray}
and for $r\rightarrow \infty$
\begin{eqnarray}
v_{1}^{l}(r) &\stackrel{\infty }{\sim }& v_{0}^{l}(r), \\
{\pep{v}}_{1}^{l}(r) &\stackrel{\infty }{\sim }& v_{0}^{l}(r).
\end{eqnarray}
The creation/annihilation operators become 
\begin{eqnarray}
\hat{a}_{0}^{\pm } &\stackrel{\infty }{\sim }& \mp \frac{i\hat{p}}{\sqrt{2m}}
-\gamma _{0}, \\
\hat{\pep{a}}_{0}^{\pm } &\stackrel{\infty }{\sim }& \mp \frac{i\hat{p}}{\sqrt{2m}}
+\gamma _{0}.
\end{eqnarray}
Using these asymptotic expressions, one find the following properties of
the mapping operators: 
\begin{eqnarray}
\langle r|\hat{\mathrm{u}}_{0}^{\pm }|r^{\prime }\rangle &\stackrel{\infty }
{\sim }&\langle r|\frac{\mp i\hat{p}/\sqrt{2m}-\gamma _{0}}
{\sqrt{\hat{p}^{2}/2m+\gamma _{0}^{2}}}|r^{\prime }\rangle, \\
\langle r|\hat{\mathrm{\pep{u}}}_{0}^{\pm }|r^{\prime }\rangle &\stackrel{\infty}
{\sim }&\delta (r-r^{\prime }).
\end{eqnarray}

\begin{figure}[htb]
\center
\includegraphics[scale=0.3]{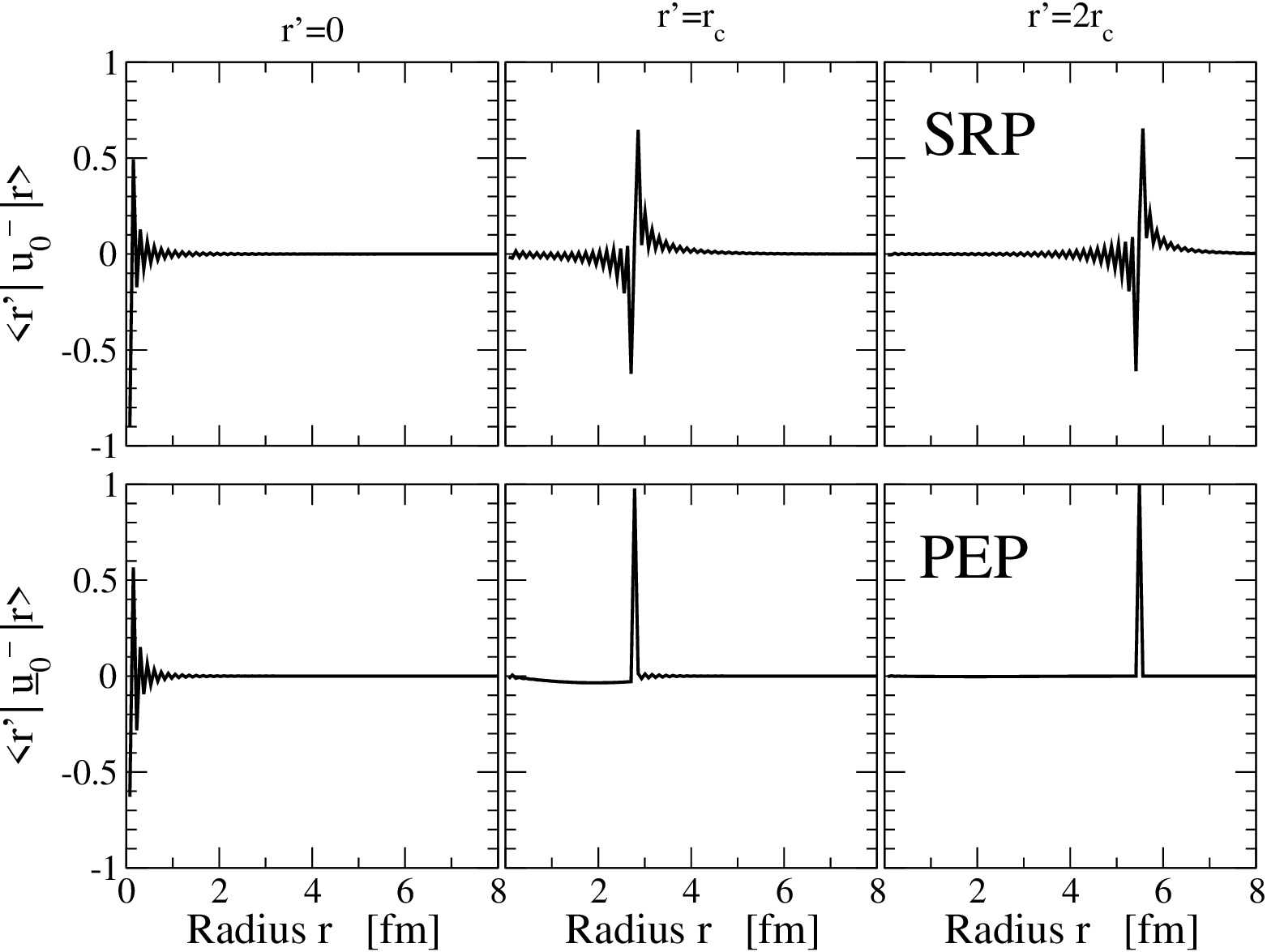}
\caption{Matrix elements of $\mathrm{u}_{0}^{-}(r,r^{\prime })$ (top, SRP) and 
$\mathrm{\pep{u}}_{0}^{-}(r,r^{\prime })$ (bottom, PEP) for several values of 
$r^{\prime}$ ($r^{\prime }=0$, $r_{c}$, $2r_{c}$) calculated for the angular 
momentum $l=0$ inside the core-halo potential.
We remind that $r_{c}$ is the radius of the $^{10}$Be core.}
\label{figu03}
\end{figure}

We present in Fig.~\ref{figu03} the matrix elements of 
$\langle r|\hat{\mathrm{u}}_{0}^{-}|r^{\prime }\rangle$ 
and $\langle r|\hat{\mathrm{\pep{u}}}_{0}^{-}|r^{\prime }\rangle $ 
as a function of $r$ for several values of 
$r^{\prime }$: $0,r_{c}$ and $2r_{c}$ where $r_{c}$ is the radius of
the $^{10}$Be core. The peaks identifies the diagonal terms. We
remark that the operator $\hat{\mathrm{u}}_{0}^{-}$ has important
off-diagonal terms (for small and large values of $r^{\prime }$) while the
operator $\hat{\mathrm{\pep{u}}}_{0}^{-}$ converges towards a delta function when 
$r^{\prime }$ increases. Hence, the restoration of the phase shift 
imposes $\hat{\mathrm{\pep{u}}}_{0}^{\pm}\sim \hat{1}$ for large
values of $r^{\prime }$, however this relation breaks down close to the core
where the off-diagonal terms become important. 


\section{Electric excitation \label{section:4}}

In this section, we shall consider dynamical properties of nuclei within the
consistent SUSY transformation we have developed in the previous sections
and discuss how it may be approximated. We will discuss the
modifications of the excitation operators, the doorway sensitivity, and
finally, we will compute some transition elements and strength associated to
monopole (E0), dipole (E1) and quadrupole (E2) electromagnetic
excitations and the associated sum rules.
 
We assume that, prior to any SUSY transformation,
the  excitation operator takes the standard multipolar form  
\begin{equation}
\hat{f}_{0}(\lambda ,L,M)=\hat{f}_0^\mathrm{rad}(\lambda) \hat{Y}_{L,M},
\label{excop}
\end{equation}
with the radial excitation operator 
$\hat{f}_0^\mathrm{rad}(\lambda)=\hat{r}^\lambda$.
We drop the coupling constant
because we are only interested in the transformation of the radial
excitation operator and the relative difference 
between the consistent SUSY transformation and its approximations. 
The E0 transition is induced by $\hat{f}_{0}(2,0,0)$ 
and the electromagnetic transitions E$\lambda $ are induced
by $\hat{f}_{0}(\lambda ,\lambda ,M)$ ($\lambda \ge 1$). The SUSY
transformation of the excitation operator is
\[
\hat{f}_{k}(\lambda ,L,M)=\hat{\mathrm{u}}_{k-1}^{+}\,
\hat{f}_{0}(\lambda,L,M)\,\hat{\mathrm{u}}_{k-1}^{-} .
\]

Introducing explicitly the angular momentum quantum numbers, the radial
excitation operator is thus given by  
\begin{equation}
\hat{f}_{k}^{l^{\prime }l}(\lambda )=\hat{\mathrm{u}}_{k-1}^{l^{\prime }+}\,
\hat{f}_{0}^\mathrm{rad}(\lambda )\,\hat{\mathrm{u}}_{k-1}^{l-}, \label{excitation}
\end{equation}
where $\hat{\mathrm{u}}_{k-1}^{l\,\pm }$ is the mapping operators 
$\hat{\mathrm{u}}_{k-1}^{\pm }$ in a subspace associated with the angular 
momentum $l$ (and $m$). The external operator $\hat{f}_{k}^{l^{\prime }l}(\lambda)$
allows transitions between different angular momentum space according to the
selection rules deduced from the relation
\begin{equation}
\langle l^{\prime }m^{\prime }|\hat{f}_{k}(\lambda ,L,M)|lm\rangle =
\hat{f}_{k}^{l^{\prime }l}(\lambda )\langle l^{\prime }m^{\prime }|
\hat{Y}_{L,M}|lm\rangle .
\end{equation}
It should be noticed that, in Eq.~\ref{excitation}  the mapping operator 
$\hat{\mathrm{u}}_{k-1}^{l\,\pm }$ on the right and on the left side of 
$\hat{f}_{0}^\mathrm{rad}(\lambda)$ may not correspond to the same angular 
momentum $l$. Moreover, while the original radial excitation operator 
$\hat{f}_{0}^\mathrm{rad}(\lambda)$ is diagonal (in the coordinate space), 
$\hat{f}_{k}^{l^{\prime}l}(\lambda )$ 
is no longer diagonal because the transformation operators 
$\hat{\mathrm{u}}_{k-1}^{l\,\pm }$ are non local. 

\subsection{Consistent excitation operator and its approximations}

In the following, we shall perform the calculation of the excitation
operator and some of the observables it induces. 
In the literature, the SUSY transformation is in general not applied to the
excitation operator. Hence, instead of calculating the matrix elements
induced by the consistent excitation operator 
$\hat{f}_{k}^{l'l}(\lambda)$, 
the authors have evaluated the matrix elements of 
$\hat{f}_{0}^\mathrm{rad}(\lambda)$ with the SUSY transformed
wave functions. We will refer to this approximation 
as the \textit{internal} approximation. 
We introduce a second approximation called the \textit{diagonal}
approximation which consists simply of neglecting the off-diagonal terms in
coordinate space of the consistent excitation operator.

As a first example we will study the $E0$ excitation of the halo neutron in
the $s$-subspace. In this subspace the core blocks one orbital (1s)
so that we have to perform a SRP or a PEP  transformation to
remove this occupied state from the halo Hilbert space and restore the phase 
shift. Of course, to be
complete we have also to remove the occupied $p$-state but since the SUSY
transformation is block-diagonal for the angular momentum quantum numbers
this does not modify the dynamics in the $s$-subspace.

\begin{figure}[htb]
\center
\includegraphics[scale=0.3]{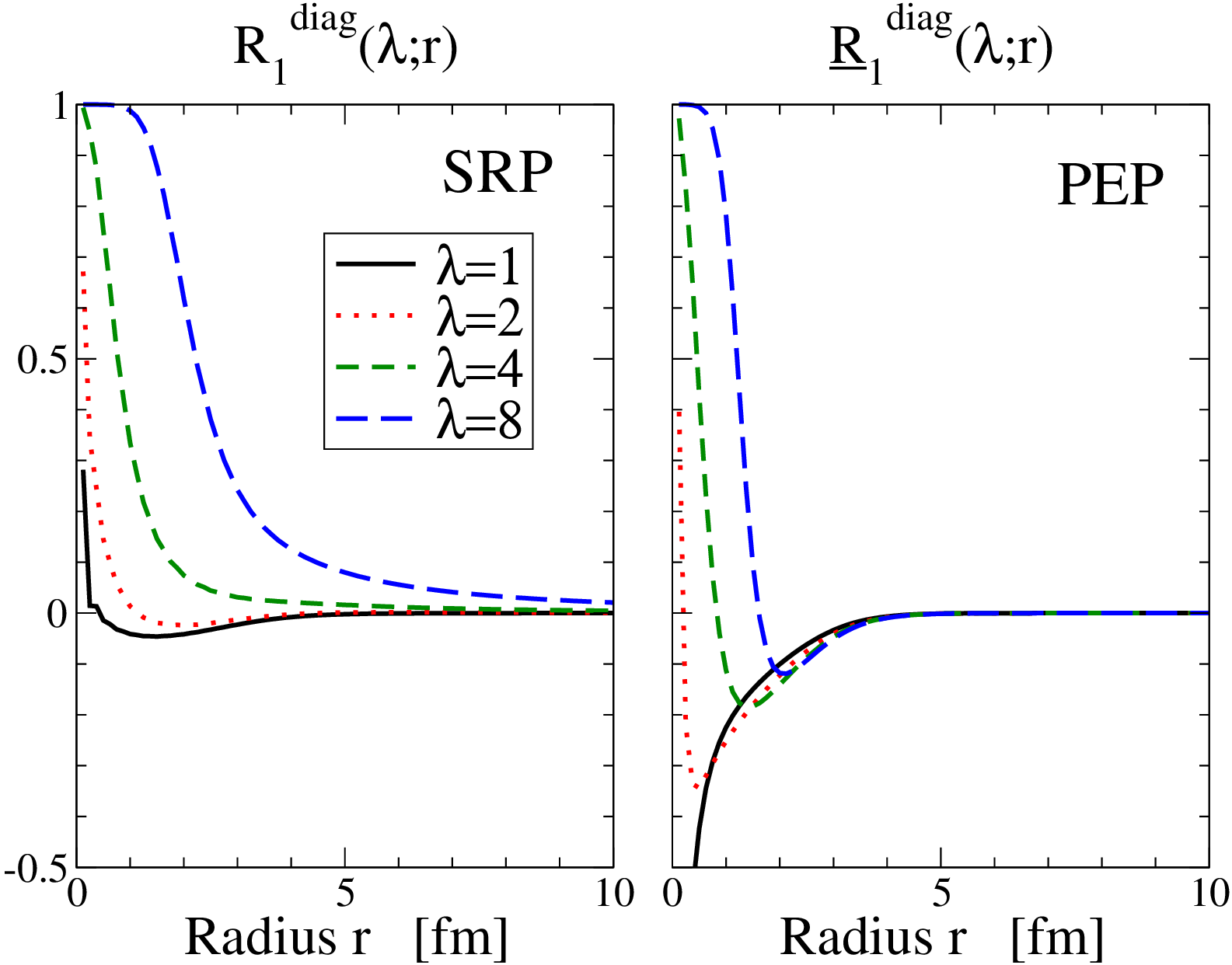}
\caption{(color online) The normalized difference of diagonal matrix elements
($R_{k}^{\mathrm{diag}}(\lambda ;r)$ for SRP transformation and 
${\pep{R}}_{k}^{\mathrm{diag}}(\lambda ;r)$ for PEP one) 
are represented for several values of 
$\lambda $ (1,2,4,8) for the electric excitation $E\lambda$ inside
the s-space.}
\label{figu04}
\end{figure}

In order to evaluate the difference between the consistent SUSY
transformation and its approximations, we define two quantities 
\begin{eqnarray}
R_{k}^{\mathrm{diag}}(\lambda;r) &=&\frac{\langle r|\Delta 
\hat{f}_{k}^{00}(\lambda)
|r\rangle }{\langle r|\hat{f}_{k}^{00}(\lambda)|r\rangle }, \\
R_{k}^{\mathrm{off}}(\lambda;r,r^{\prime }) &=&\frac{\langle r|
\hat{f}_{k}^{00}(\lambda)
|r^{\prime }\rangle }{\langle r|\hat{f}_{k}^{00}(\lambda)|r\rangle },
\end{eqnarray}
where 
\[
\Delta \hat{f}_{k}^{l'l}(\lambda)=\hat{f}_{k}^{l'l}(\lambda)-
\hat{f}_{0}^\mathrm{rad}(\lambda) 
\]
is the difference between the excitation operator consistently transformed 
$\hat{f}_{k}^{l'l}(\lambda)$ and the original excitation operator 
$\hat{f}_{0}^\mathrm{rad}(\lambda)$. 
The ratio $R_{k}^{\mathrm{diag}}(\lambda;r)$ evaluates the difference between the
diagonal part of the consistent excitation operator and the original
operator, normalized to the value of the diagonal part of the consistent
operator. It gives an evaluation of the approximation for the diagonal part
of the excitation operator. Fig.~\ref{figu04} shows the ratio 
$R_{k}^{\mathrm{diag}}(\lambda;r)$ and $\pep{R}_{k}^{\mathrm{diag}}(\lambda;r)$ 
for $\lambda $=1, 2, 4, 8 in the case of the 1s SRP and PEP transformation
respectively. We remark that for large radii $r$, 
the diagonal part of the excitation operator is close to the original
one ($R_{k}^{\mathrm{diag}}(\lambda;r)\sim 0$). This is a consequence of the
asymptotic properties of the mapping operator as it has been discussed in
the section \ref{pot:halo}. 
For small radii $r$, the diagonal
part of the excitation operator is strongly modified, the ratio 
$R_{k}^{\mathrm{diag}}(\lambda;r)\sim 1$  revealing that 
$\langle r|\hat{f}_k^\mathrm{rad}|r\rangle \gg \langle r|\hat{f}_0^\mathrm{rad}|r\rangle$. 
This difference persists through a large range of radial coordinates. 
This range increases with $\lambda $, it is wider for SRP transformation
compared with PEP transformations.

\begin{figure}[htb]
\center
\includegraphics[scale=0.3]{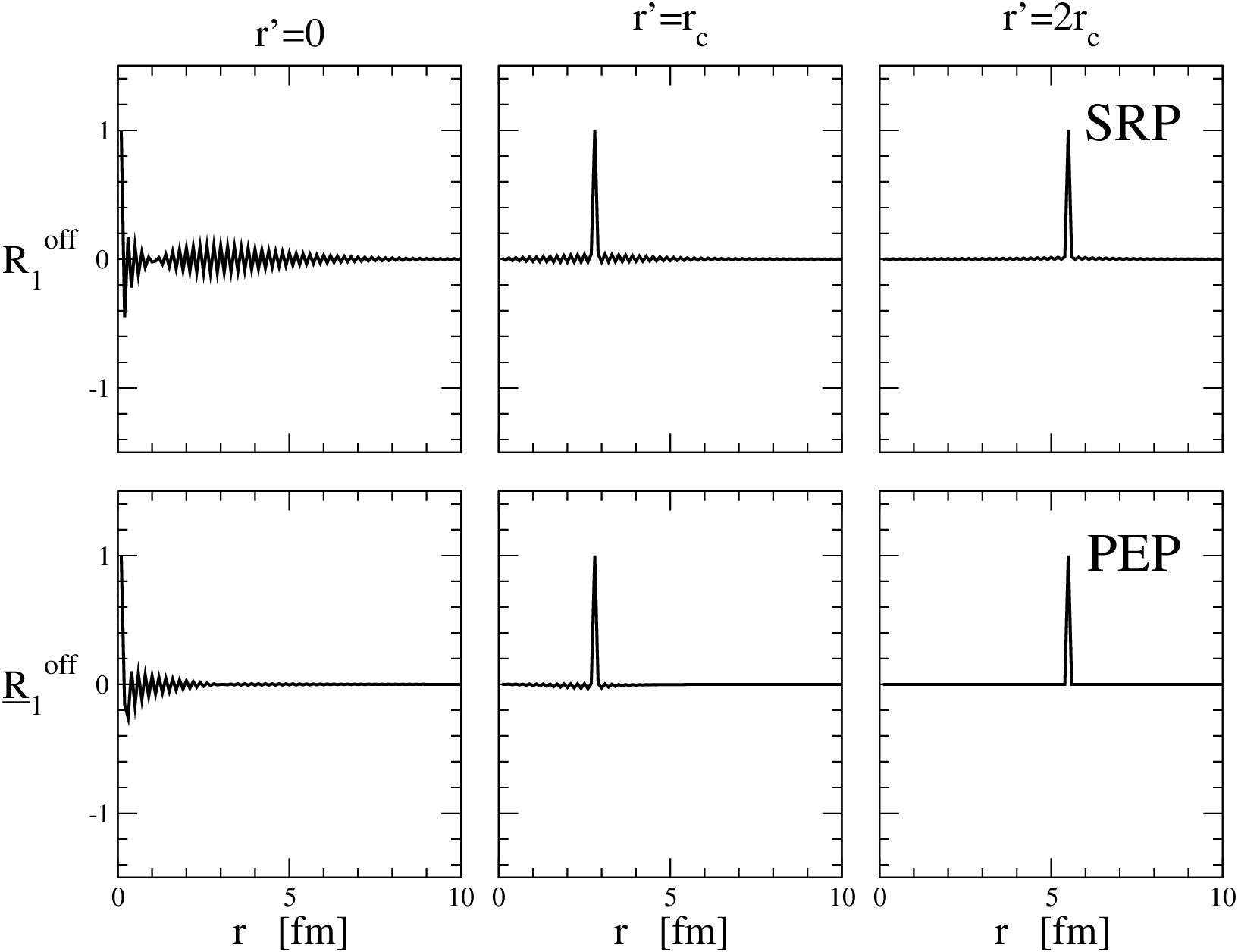}
\caption{The normalized off-diagonal matrix elements
($R_{k}^{\mathrm{off}}(\lambda =4;r,r^{\prime })$ and 
${\pep{R}}_{k}^{\mathrm{off}}(\lambda =4;r,r^{\prime })$) 
are represented as a function of $r$ for several values of r' 
(0, $r_{c}$, $2r_{c}$) for the electric excitation 
$E\lambda$ inside the s-space. }
\label{figu05}
\end{figure}

On the other hand, the ratio $R_{k}^{\mathrm{off}}(\lambda;r,r^{\prime })$ evaluates
the amplitude of the off-diagonal terms in coordinate space normalized to
the diagonal term of the consistent excitation operator. In Fig.~\ref{figu05},
we represent the ratio $R_{k}^{\mathrm{off}}(\lambda=4;r,r^{\prime })$ and
$\pep{R}_{k}^{\mathrm{off}}(\lambda=4;r,r^{\prime })$ for the SRP
and PEP transformation respectively, and for three values of $r^{\prime }$: 0, 
$r_{c}$ and $2r_{c}$. For SRP transformation, off diagonal terms are
important for small $r^{\prime }$ and decrease in relative magnitude while 
$r^{\prime }$ increases. Off-diagonal terms are non zero in a wide range and
we will show in the next paragraphs that they can have a more important
effect on observables than the diagonal terms. For the PEP transformation
the off diagonal terms are smaller and become negligible for intermediate
and large $r^{\prime }$ ($\ge r_{c}$) as required by the restoration of the
asymptotic behavior.

In all the cases presented here, the consistent excitation operator is
different from the original one in the space region inside the core
potential. Hence, from this observation, we can expect that there will be
important effects induced by the consistent calculation if and only if the
calculated observable is sensitive to the space region inside the core
potential.

\subsection{Transformation of the doorway state}

We want now to evaluate both contributions of the diagonal and off-diagonal
excitation operators on the matrix elements. For that, we introduce the
doorway state defined as 
\[
|\delta \varphi _{k}(\lambda,l\rightarrow l' )\rangle =
\hat{f}_{k}^{l'l}(\lambda)|\varphi_{k}^l\rangle. 
\]
In the following, we have chosen for $|\varphi_{k}^l\rangle$ the ground state of 
$\hat{h}_k^l$.

\begin{figure}[htb]
\center
\includegraphics[scale=0.3]{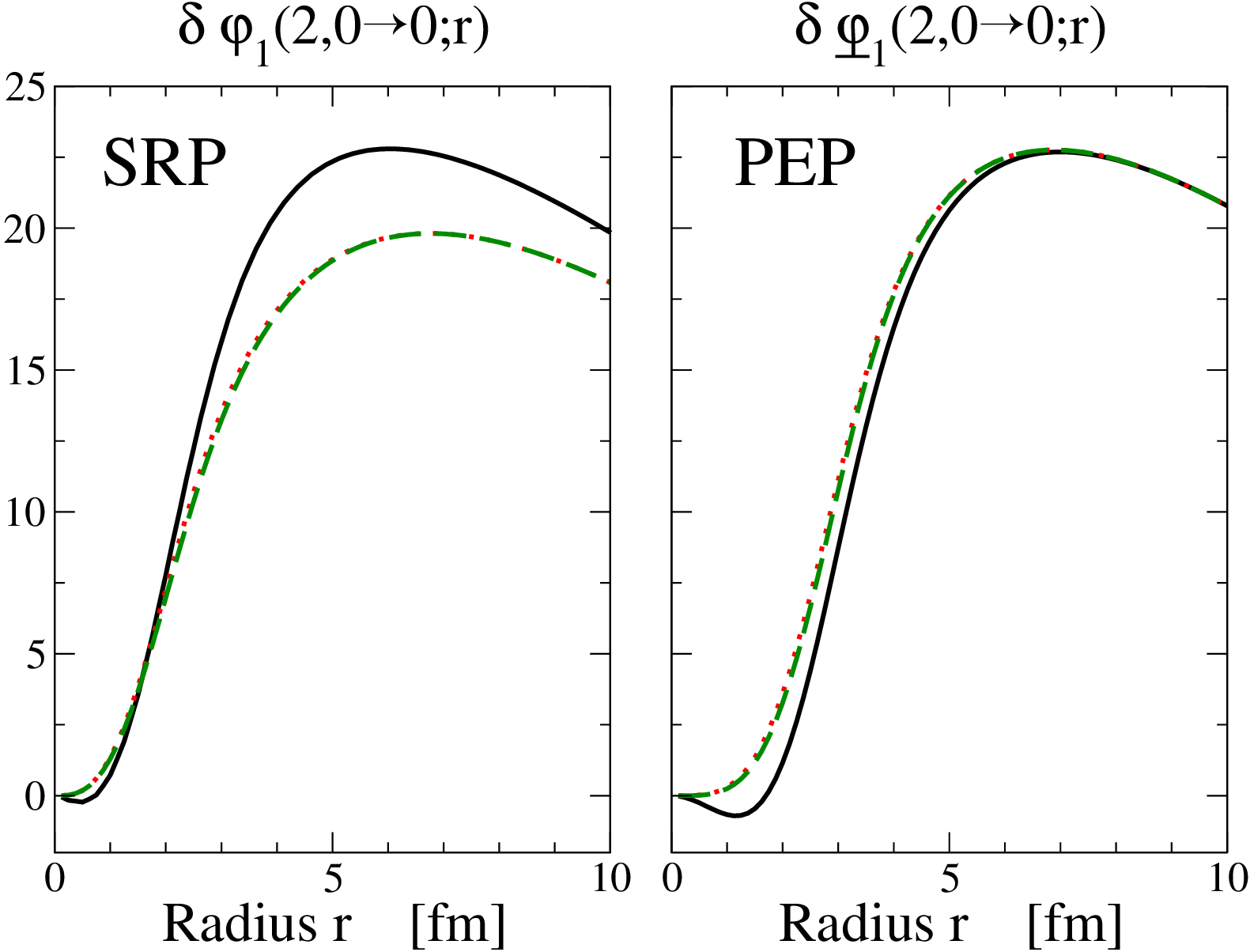}
\caption{(color online) The doorway state 
$\delta \varphi_{1}(\lambda=2, 0\rightarrow 0;r)$ and
$\delta \pep{\varphi}_{1}(\lambda=2, 0\rightarrow 0;r)$ 
of electric excitations for SRP and PEP transformations respectively. 
The solid line stands for the consistent excitation operator, 
dotted line for the internal approximation and dashed line 
for the diagonal approximation.}
\label{figu06}
\end{figure}

We represent in Fig.~\ref{figu06} the doorway state 
$\delta \varphi_{1}(\lambda=2,0\rightarrow 0;r)$ and
$\delta \pep{\varphi}_{1}(\lambda=2,0\rightarrow 0;r)$ 
for the SRP and PEP 
transformations respectively. The solid line stands for
the consistent excitation operator, dotted line for the \textit{internal}
approximation and dashed line for the \textit{diagonal} approximation. We
remark that the \textit{internal} approximation and the \textit{diagonal}
approximation are indistinguishable. This shows that off-diagonal terms are
the most important sources of modification of the excitation operator.

Moreover, the consistent doorway state changes sign while the two
approximations remains positive. This affects the node structure of the
wave function and may induce strong modifications for forbidden transition as
we will see in the next paragraph.

\subsection{Single particle reduced transition probability}

The single-particle reduced transition probabilities are defined as 
\cite{Bor69} 
\begin{eqnarray}
B_{k}(E0,i\rightarrow f) =\langle \varphi _{k}^{f}|
\hat{f}_{k}^{l_{f}\,l_{i}}(2)|\varphi _{k}^{i}\rangle
\end{eqnarray}
with $l_{f}=l_{i}$ and as
\begin{eqnarray}
B_{k}(E\lambda ,i\rightarrow f) &=&\langle \varphi _{k}^{f}|
\hat{f}_{k}^{l_{f}\,l_{i}}(\lambda )|\varphi _{k}^{i}\rangle
\end{eqnarray}
with $\left| l_{i}-\lambda \right| \leq l_{f}\leq l_{i}+\lambda$ and
for $\lambda \ge 1$. In order to simplify the notations, the states $i$ and 
$j$ are labeled according to the original space prior to any transformation.
In the halo case, developed above, all the final states are in the continuum
so it will not be possible to use directly these definitions.  In the next
section we will introduce the strength function, a more general way to look
at transition probabilities which is suitable for the case of excitation
towards continuum and which can thus be used in the halo case. To get 
results for the transition probabilities between discreet states, in the
present section, we will restrict the discussion to
the harmonic oscillator model (see section \ref{pot:ho}). 
To simplify the discussion, we
will consider that the nucleons of the core only occupy the $1s$ 
orbital 
and we will 
study the excitation of an additional neutron in the $2s$ orbital. We
have computed numerically the reduced matrix elements ${\pep{B}}_{1}$(E0) and 
${\pep{B}}_{1}$(E1) for the PEP transformation. 
The results are presented respectively in the tables 
\ref{table:1} and \ref{table:2}. In the harmonic oscillator, 
due to selection rules, from the
2s the monopole operator $\hat{r}^{2}$ can induce transition only towards the 3s.
In table \ref{table:1}, the first line shows the result of the matrix
elements ($B$) induced by the consistent excitation operator. As expected, the
forbidden transition are zero within the numerical uncertainty indicated in
parenthesis.

The matrix elements, $\widetilde{B}$, 
induced by the \textit{internal} excitation operator are
showed in the second line of table \ref{table:1}. 
For allowed transitions, the \textit{internal}  approximation
modifies by about 20\% the exact matrix element, but the main effect of this
approximation is that it induces forbidden transitions from 2s to 4s-7s
states.

On the other hand, in the case of E1 electromagnetic transition, the
selection rules of dipole transitions in the harmonic oscillator allow
transition from 2s states to 1p and 2p states. The same phenomenon is
observed in table \ref{table:1} and table \ref{table:2}: the 
\textit{internal} approximation produces spurious excitation of forbidden  states.
Moreover, for the allowed transition the error goes up to more then a factor $2$.

\begin{table}[htb]
\begin{tabular}{c|ccccc}
\hline
f & 3s & 4s & 5s & 6s & 7s \\ \hline\\[-3mm]
${\pep{B}}_1$(E0) & 1.1$\times$10$^{4}$ & o(10$^{-4}$) & o(10$^{-8}$) & o(10$^{-7}$)
& o(10$^{-7}$) \\[0.1cm]
$\widetilde{{\pep{B}}}_1$(E0) & 9.2$\times$10$^{3}$ & 1.5$\times$10$^{1}$ & 1.3 & 7.1$%
\times$10$^{-2}$ & 1.4$\times$10$^{-4}$ \\ 
\hline
\end{tabular}
\caption{${\pep{B}}_1$(E0, 2s$\rightarrow$ f) for the PEP transformed
harmonic oscillator potential. 
$\widetilde{\pep{B}}_1$(E0, 2s$\rightarrow$ f) is the matrix element induced by the 
\textit{internal} approximation of the excitation operator.}
\label{table:1}
\end{table}

\begin{table}[htb]
\begin{tabular}{c|ccccc}
\hline
f & 1p & 2p & 3p & 4p & 5p \\ \hline\\[-3mm]
${\pep{B}}_1$(E1) & 5.0$\times$10$^{2}$ & 1.2$\times$10$^{3}$ & o(10$^{-6}$) & o(10$%
^{-7}$) & o(10$^{-6}$) \\[0.1cm]
$\widetilde{{\pep{B}}}_1$(E1) & 1.2$\times$10$^{3}$ & 8.0$\times$10$^{2}$ & 9.5 & 1.2
& 7.8$\times$10$^{-2}$ \\ 
\hline
\end{tabular}
\caption{$\pep{B}_1$(E1, 2s$\rightarrow$ f) for the PEP transformed harmonic oscillator potential. 
$\widetilde{\pep{B}}_1$(E1, 2s$\rightarrow$ f) is the matrix element induced by the 
\textit{internal} approximation of the excitation operator.}
\label{table:2}
\end{table}

\subsection{Strength}

The very important discrepancy between the internal and the
complete SUSY observed in the case of the harmonic oscillator might be a
peculiarity due to the symmetry of this model. 
Let us thus come back to the physical case of the $^{11}$Be halo nuclei 
for which transitions between orbitals belonging to the same l-space
are all allowed. 
Now we should remove the  $1s$ (-25 MeV) and $1p$ (-12 MeV)  
orbitals occupied by the core neutrons so that only one bound state 
($2s$) is available for the halo neutron. 
The excitations can only promote the halo neutron to the continuum.
Hereafter, the eigenstates and the continuum states will be obtained
from the diagonalization of the Hamiltonian inside a box going up
to 50~fm with 400 points.

In order to discuss transition towards the continuum, we introduce the
strength: 
\begin{eqnarray}
S_{k}(\mathrm{E}\lambda ,i,\omega ) &=&\sum_{f}|B_{k}
(\mathrm{E}\lambda, i\rightarrow f)|^{2}\delta (\omega -E_{f}+E_{i}), 
\nonumber\\
\end{eqnarray}
where $i$ is the initial state, here the $2s$ orbital, and the final states 
$f$ are the continuum states of the box.
The single particle energies are $E_{i}$ and $E_{f}$ respectively.
Since we perform the calculation in a box the continuum is discreetized. In
order to obtain a smooth strength function one often smoothes the obtained
results with a Gaussian or a Lorentzian function. In this paper we 
will do both.

\begin{figure}[htb]
\center
\includegraphics[scale=0.3]{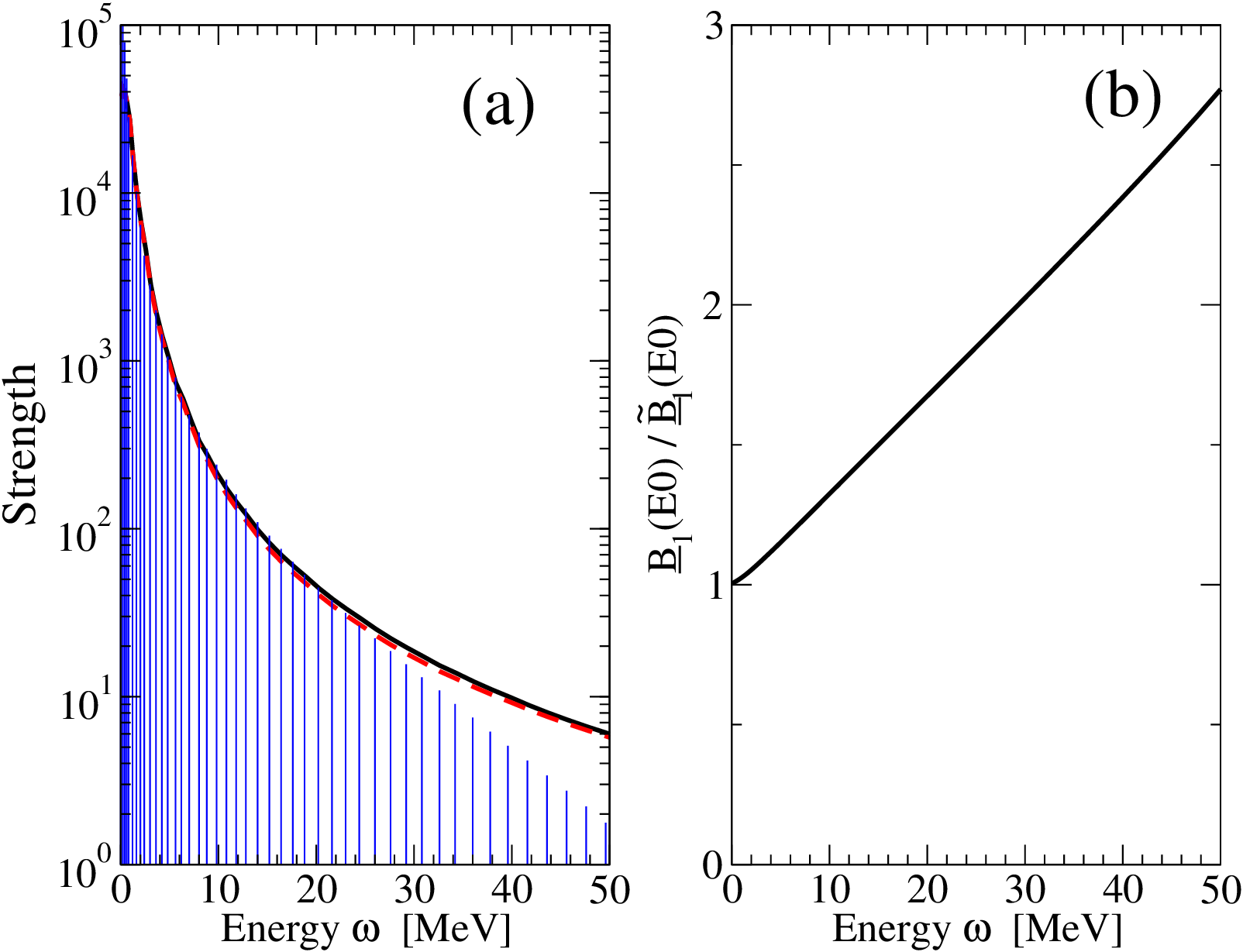}
\caption{(color online) For monopole excitation: Left (a) the bars are the reduced
transition probabilities ${\pep{B}}_{1}$(E0) for the continuum states discreetized in
the considered box, the solid line is the strength function resulting
from a Lorentzian smoothing ($\Gamma $=500 keV) and the dashed line
is the result of the \textit{internal} approximation. 
In part (b), we show the ratio ${\pep{B}}_{1}$(E0)/$\widetilde{\pep{B}}_{1}$(E0).}
\label{figu07}
\end{figure}

\begin{figure}[htb]
\center
\includegraphics[scale=0.3]{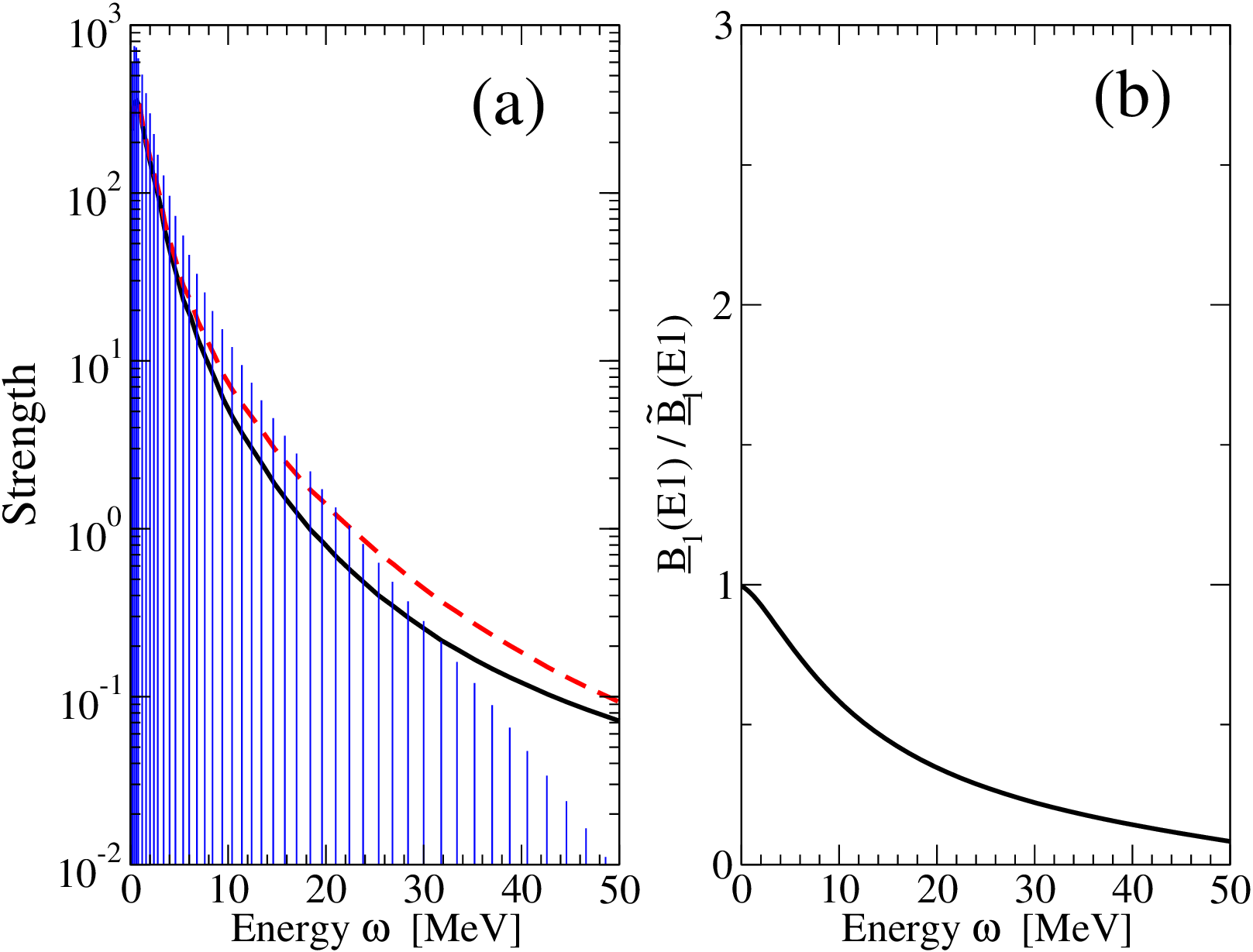}
\caption{(color online) Same as Fig.~\ref{figu07} for dipole transitions.}
\label{figu08}
\end{figure}

\begin{figure}[htb]
\center
\includegraphics[scale=0.3]{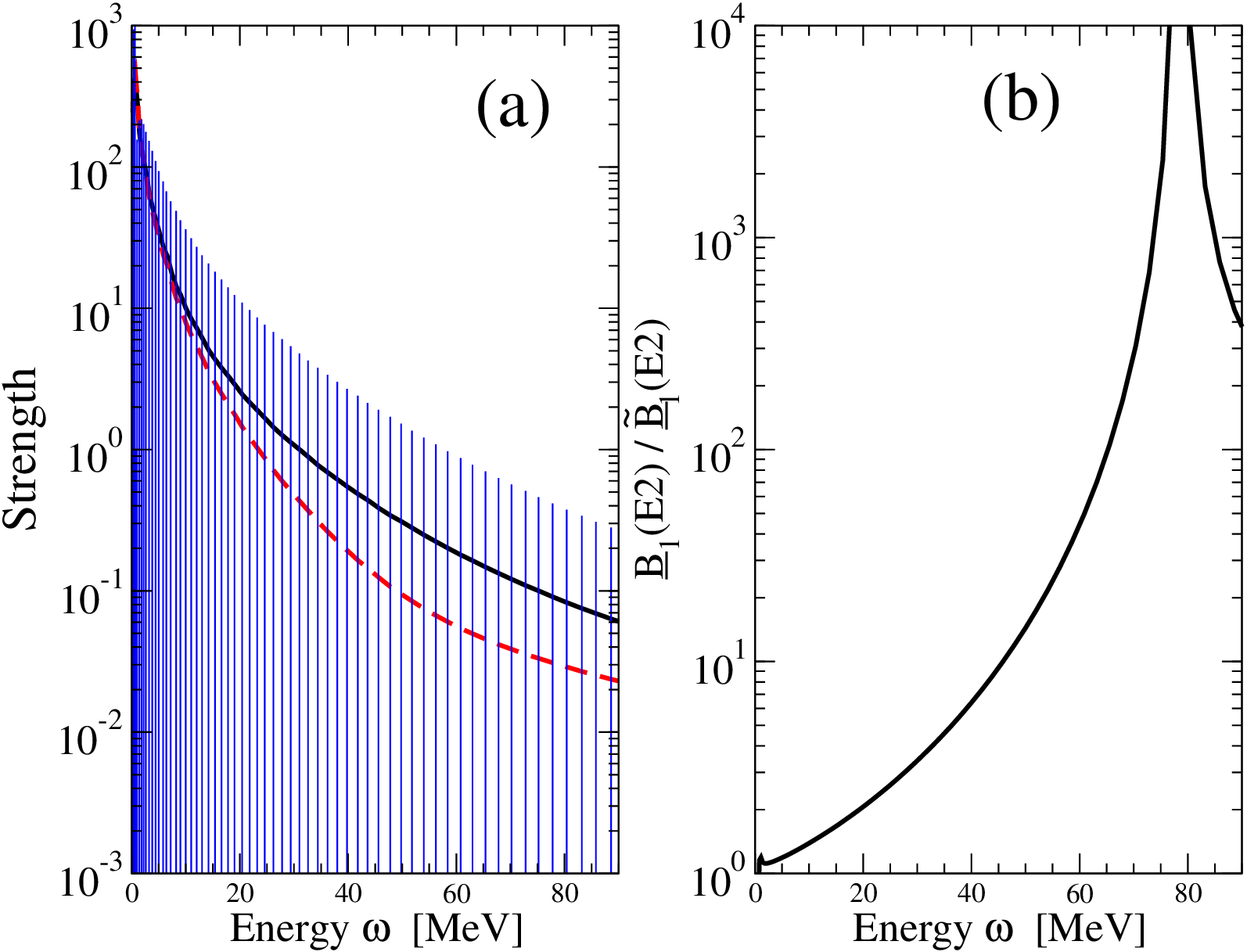}
\caption{(color online) Same as Fig.~\ref{figu07} for quadrupole transitions.}
\label{figu09}
\end{figure}

\begin{figure}[htb]
\center
\includegraphics[scale=0.3]{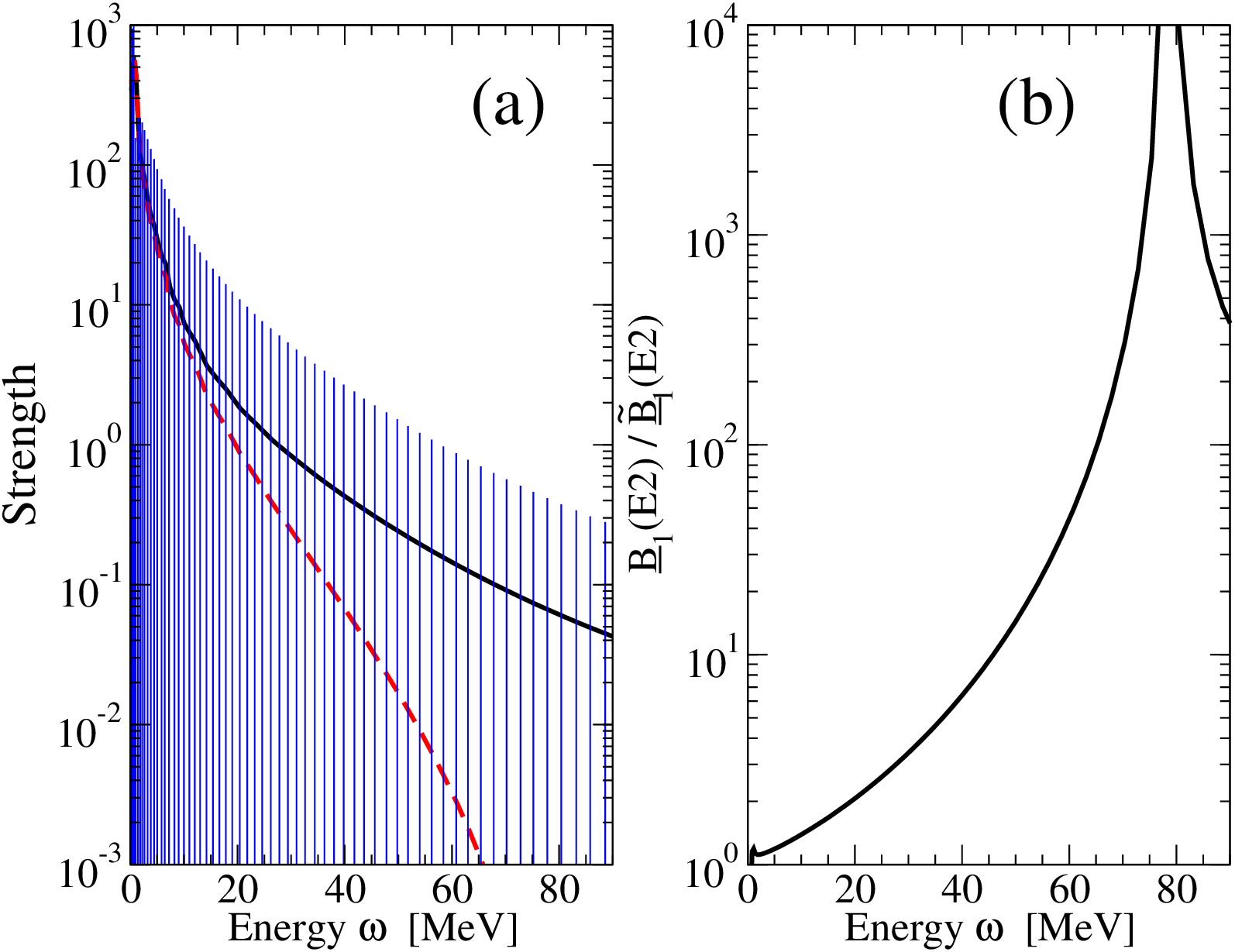}
\caption{(color online) Same as Fig.~\ref{figu09} with Gaussian smoothing.}
\label{figu10}
\end{figure}

Strengths for PEP transformations for E0, E1 and E2 transitions are
presented in the parts (a) of Fig.~\ref{figu07}-\ref{figu08}-\ref{figu09}
with a Lorentzian smoothing ($\Gamma =500$ keV) and Fig.~\ref{figu10} with a
Gaussian smoothing. The part (b) of each figure gives directly the ratio 
${\pep{B}}_{1}$(E$\lambda$ )/$\widetilde{\pep{B}}_{1}$(E$\lambda$ ) computed for individual
states.

For the monopole mode, using a Lorentzian smoothing the consistent strength
and the internal strength appears to be very similar (see 
Fig.~\ref{figu07}-a) even if the ratio ${\pep{B}}_{1}$(E0)/$\widetilde{\pep{B}}_{1}$(E0) 
computed for individual states (see part (b) of the figure) is very 
different from $1$ for large values of the final excitation energy $E_{f}$. 

The dipole excitations  connect the states of two different l-subspace. On
the smoothed ${\pep{S}}_1$(E1) strength (see Fig.~\ref{figu08}-a) we only observe
a small over estimation of the strength for large values of $E_{f}$ but
again the effect seems much larger on the ratio 
${\pep{B}}_{1}$(E1)/$\widetilde{\pep{B}}_{1}$(E1). 
In Fig.~\ref{figu09}, we represent the strength ${\pep{S}}_1$(E2). In
the l=2 subspace, there are no core states and consequently, the SUSY
transformation is in fact the unity. The smoothed strength appears to be
only slightly under-estimated by the internal approximation again in
contradiction with the ratio ${\pep{B}}_{1}$(E2)/$\widetilde{\pep{B}}_{1}$(E2) which
exhibits a strong discrepancy. 

In order to solve the contradiction 
we have first studied the role of the smoothing. We have found that the
situation is different with a Gaussian smoothing as illustrated in 
Fig.~\ref{figu10}. This difference is due to the difference in the tails of 
the two smoothing functions: the long tails of the Lorentzian function
associated with the low energy states which have a large ${\pep{B}}_{1}$(E0)
dominate even at large energy when a Lorentzian form factor is
used. Indeed, since the difference between the two calculations is small for
these dominating states the final Lorentzian-smoothed strengths for the two
calculations are rather close in contradiction with the direct ratio of
individual excitation probabilities or the results of the Gaussian
smoothing.

To avoid the ambiguity of the smoothing method we have studied the 
continuum limit by a direct scaling of the numerical box size  
(we have also tested the role of the mesh size).
We have observed that the ratio ${\pep{B}}_{1}$/$\widetilde{\pep{B}}_{1}$ computed for each
individual state does not change shape going to the continuum limit while the
smoothed strengths vary and exhibit a strong dependence into the smoothing
functional and parameters. Therefore, the ratio 
${\pep{B}}_{1}$(E0)/$\widetilde{\pep{B}}_{1}$(E0)  provides in fact the correct 
continuum limit and the large 
observed discrepancy at high energy between the internal and the complete
SUSY transformation is the physical one.  This is even better illustrated by
considering integrated effects like effects on the sum rules.

\subsection{Sum rules}

By integrating the strength over the energy, we can define different sum
rules $m_{t}$  
\begin{equation}
m_{t}(E\lambda )\equiv \int \!\!\!d\omega \,\,\omega ^{t}S(E\lambda
,i=0,\omega ),
\end{equation}
where $t$ is the weight of the energy. 
In the frozen core approximation, the sum rules $m_0$ and $m_1$ 
can be obtained from the halo wave function according to 
\begin{eqnarray}
m_{0} &=&\langle \Phi _{\mathrm{v}}|\hat{f}_{0}
\hat{P}_{\mathrm{v}}\hat{f}_{0}|\Phi _{\mathrm{v}}\rangle 
-\left(\langle \Phi _{\mathrm{v}}|\hat{f}_{0}|\Phi _{\mathrm{v}}\rangle 
\right)^{2},
\\
m_{1} &=&\frac{1}{2}\langle \Phi _{\mathrm{v}}|[\hat{f}_{0}
\hat{P}_{\mathrm{v}},[\hat{h}_{0}^{l}\hat{P}_{\mathrm{v}},\hat{f}_{0}
\hat{P}_{\mathrm{v}}]]|\Phi _{\mathrm{v}}\rangle.
\end{eqnarray}
where $\hat{f}_{0}$ is the excitation operator defined by
Eq.~\ref{excop}.
We define the \textit{internal} approximation for the sum rule as 
$\widetilde{m}_{t}$ for which the projector $P_{\mathrm{v}}$ have been
removed. The Pauli principle is no longer respected. In order to estimate
the error induced in the calculation of $\widetilde{m}_{t}$ compared to 
$m_{t}$, we have estimated the ratios $(m_{t}-\widetilde{m}_{t})/m_{t}$
and we have found the results presented in Table~\ref{table:3}. The relative
error induced by the internal approximation increases with the weight. This
result is compatible with the results presented in Fig.~\ref{figu07} and
Fig.~\ref{figu08}: when the weight increases, the contribution of large
energy increases and the discrepancies between the consistent SUSY and its
internal approximation increase also.

\begin{table}[htb]
\begin{tabular}{c|ccc}
\hline\\[-0.3cm]
& $(\pep{m}_0-\widetilde{pep{m}}_0)/\pep{m}_0$ & $(\pep{m}_1-\widetilde{\pep{m}}_1)/\pep{m}_1$ & 
$({\pep{m}}_2-\widetilde{\pep{m}}_2)/{\pep{m}}_2$ \\[1mm]
\hline\\[-0.3cm]
$E0$ & 1.4\% & 3.9\% & 17.4\% \\[1mm]
$E1$ & -6.8\% & -33.3\% & -93.1\% \\ \hline
\end{tabular}
\caption{This table presents the relative error induced in the calculation
of the energy weighted sum rules by the \textit{internal} approximation of
PEP transformations.}
\label{table:3}
\end{table}

\section{Response to a Gaussian excitation \label{section:5}}

In the previous section, we have shown that the PEP transformation modifies
essentially the external excitation operator in the space region located
inside the core potential. The electric operators 
$\hat{r}^{\lambda}\hat{Y}_{L M}$ 
studied in the previous section which can be seen as a multiple
expansion of a Coulomb field far from the nucleus or as the low momentum
transfer limit of a plane wave scattering are strong far from the nucleus.
Hence, the effect of the PEP transformation on the excitation process has
been found to not be too large. However, this is not always the case and in
particular nuclear scattering and/or large momentum transfer reaction
correspond to much shorter distances. In order to study the effect of the
PEP transformation on such kind of scattering, in this section, we study the
response to a Gaussian excitation which can strongly overlap with the core
potential. Gaussian potential can be induced by an external nuclear
potential as well as a residual two-body interaction between particle in the
halo. In a similar spirit of the previous section, we will not study a
specific process but the response to a one body Gaussian potential centered
around $\mathbf{r}_{0}$ defined as 
\[
g_{0}(\mathbf{r})=-\frac{g_{0}}{\left( \sqrt{\pi }\mu \right) ^{3}}\exp
\left( -\frac{(\mathbf{r}-\mathbf{r}_{0})^2}{\mu^2} \right) ,
\]
where the norm of the interaction is $g_{0}$=450 MeV.fm$^{3}$ and its range
is $\mu$=2 fm. The SUSY transformation of this potential is 
\[
\hat{g}_{k}^{l'l}=\hat{\mathrm{u}}_{k-1}^{l'\, -}\,\hat{g}_{0}\,
\hat{\mathrm{u}}_{k-1}^{l\, +}.
\]
Similarly to the previous section, we define two quantities that measure the
modification of the SUSY transformation on the diagonal and off-diagonal
terms of the Gaussian potential: 
\begin{eqnarray}
Q_{k}^{\mathrm{diag}}(r) &=&\frac{\langle r|\hat{g}_{k}^{00}-
\hat{g}_{0}^{00}|r\rangle }{\langle r|\hat{g}_{k}|r\rangle }, \\
Q_{k}^{\mathrm{off}}(r,r^{\prime }) &=&\frac{\langle r|
\hat{g}_{k}^{00}|r^{\prime }\rangle }{\langle r|\hat{g}_{k}^{00}|r\rangle }.
\end{eqnarray}

\begin{figure}[htb]
\center
\includegraphics[scale=0.3]{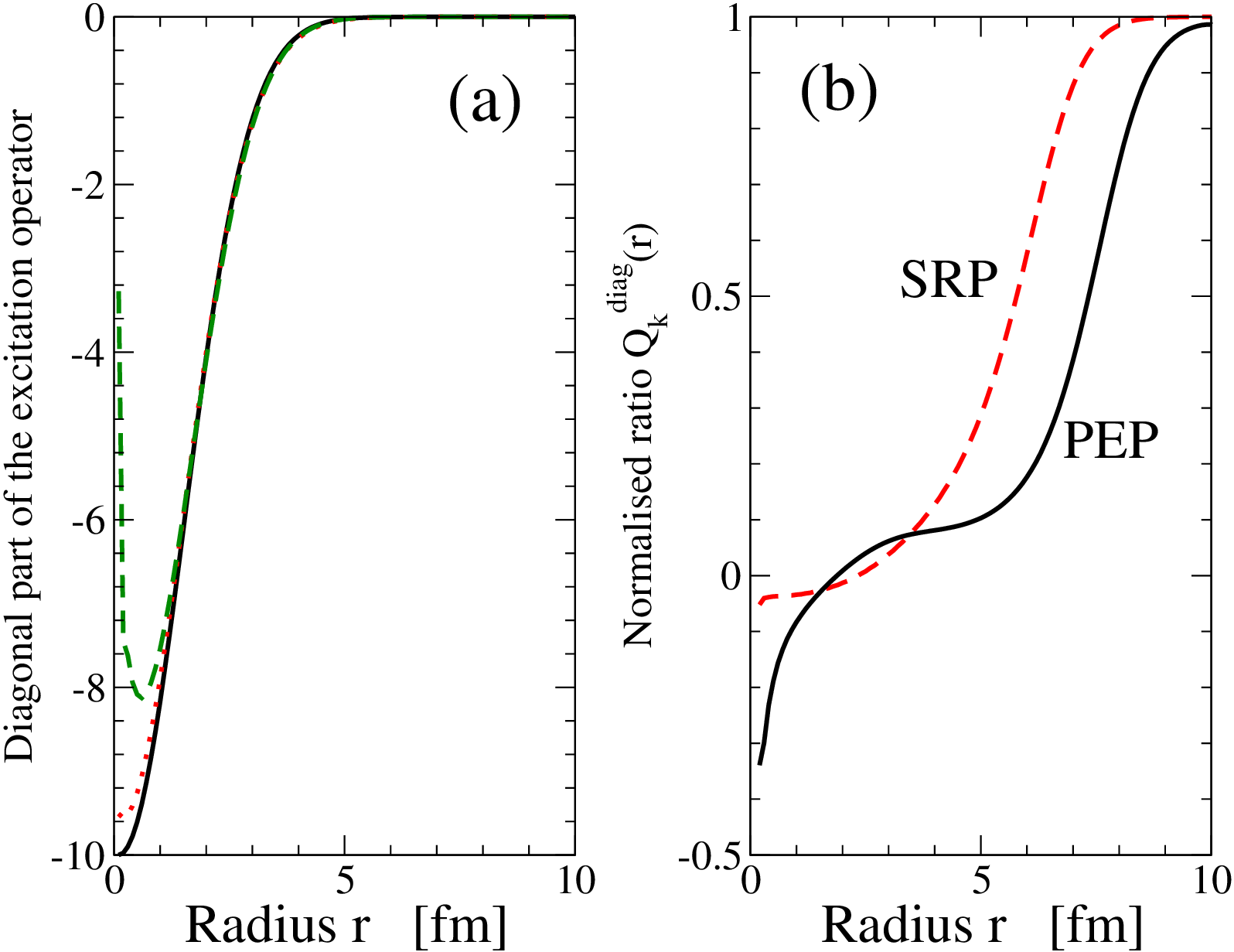}
\caption{(color online) Left (a) the initial Gaussian potential($l=0$) (solid line), SRP
(dotted line) and PEP (dashed line) potentials. In part(b), we show the
ratio $Q_{1}^{\mathrm{diag}}(r)$ for SRP (dashed) and PEP (dotted)
potentials. }
\label{figu11}
\end{figure}

In Fig.~\ref{figu11}, we fix $r_0=0$ and we represent (a) the diagonal 
part of $g_{1}$ (dotted line) and ${\pep{g}}_{1}$ (dashed line) 
compared to the original potential $g_{0}$ (solid line),
and (b) the ratio $Q_{k}^{\mathrm{diag}}(r)$ for the SRP transformation
(dashed line) and PEP transformation (solid line). In the very central
region, the PEP transformations modify the potential by a factor 30\%.
At large distance $r$, because of the Gaussian shape centered on zero
of $g_0(r')$,  an important relative weight is given to small radii 
in the summation
$\langle r|\hat{g}_k|r\rangle =\int dr' \langle r|\mathrm{\hat{u}}^- 
|r'\rangle g_0(r')\langle r'|\mathrm{\hat{u}} |r\rangle$. 
The result of this effect is that the range of 
$\langle r|\hat{g}_k|r\rangle$ is slightly increased compared to 
$g_0(r')$ and because
of the exponential behavior of the excitation operator this is 
enough to make the ratio  
$Q_{k}^{\mathrm{diag}}(r)\sim 1$.

\begin{figure}[htb]
\center
\includegraphics[scale=0.3]{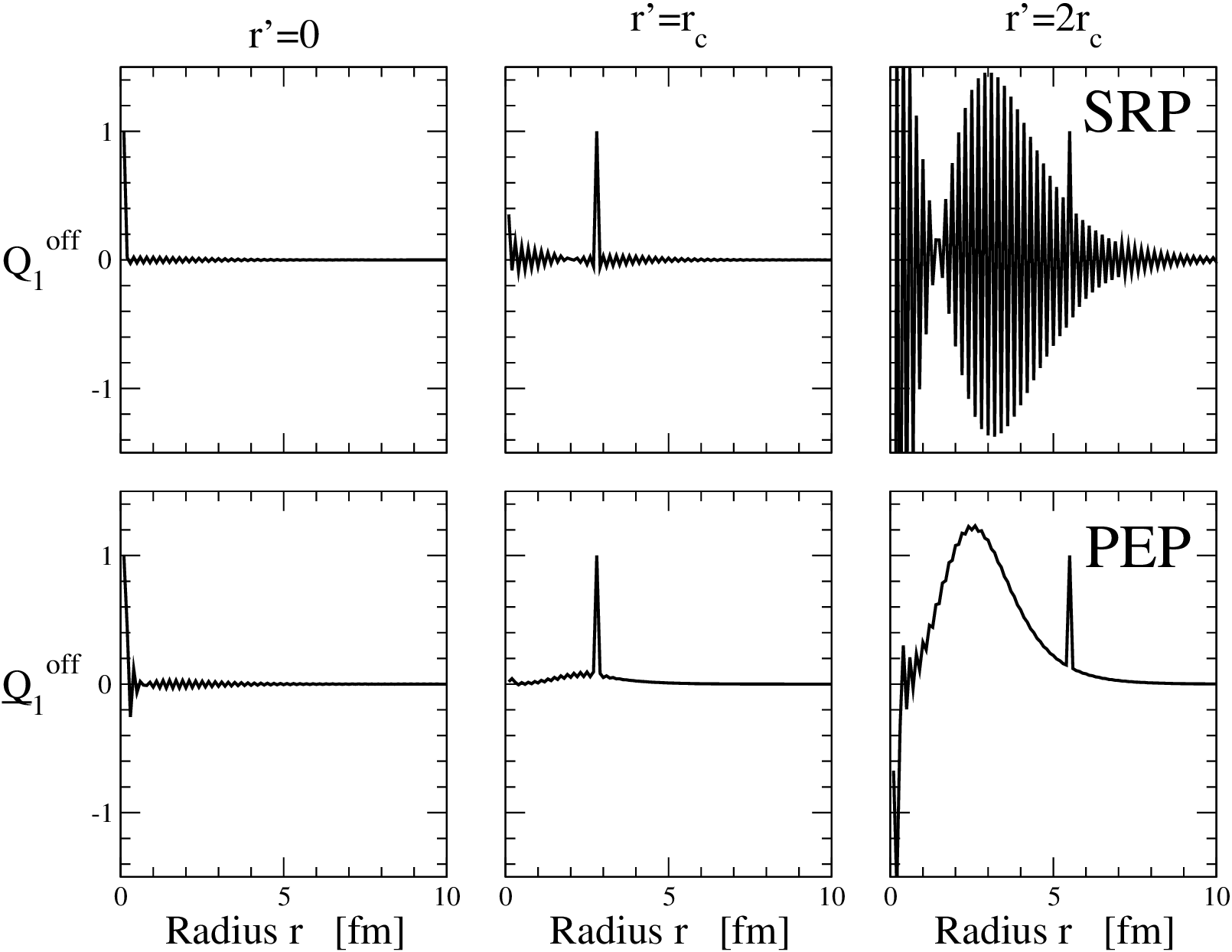}
\caption{The ratios $Q_{1}^{\mathrm{off}}(r,r^{\prime })$ in the $l=0$
channel as a function of $r$ for several values of $r^{\prime }$ ($0$, 
$r_{c} $ and $2r_{c}$) in the case of a Gaussian potential. }
\label{figu12}
\end{figure}

In Fig.~\ref{figu12}, we show the ratio $Q_{1}^{\mathrm{off}}(r,r^{\prime })$
for the SRP transformation (upper panel) and the PEP transformation (lower
panel) and for three different values of $r^{\prime }$: 0, $r_{c}$ and $2r_{c}$.
For values of $r^{\prime }$ inside the potential, off-diagonal terms are small
compared to the diagonal term but they are spread over a large range of
coordinates, and their integrated effect can counter balance their small
values. Outside the potential, the off-diagonal terms become very important
and even larger than the diagonal term for both SRP and PEP transformations.

Both diagonal and off-diagonal terms have an effect on the particle wave
function that can be estimated with the doorway state 
\[
|\delta \varphi _{k}(l\rightarrow l')\rangle =\hat{g}_{k}^{l'l}
|\varphi_k^l\rangle. 
\]
In the following, we have chosen for $|\varphi_{k}^l\rangle$ the ground state of 
$\hat{h}_k^l$.

\begin{figure}[htb]
\center
\includegraphics[scale=0.3]{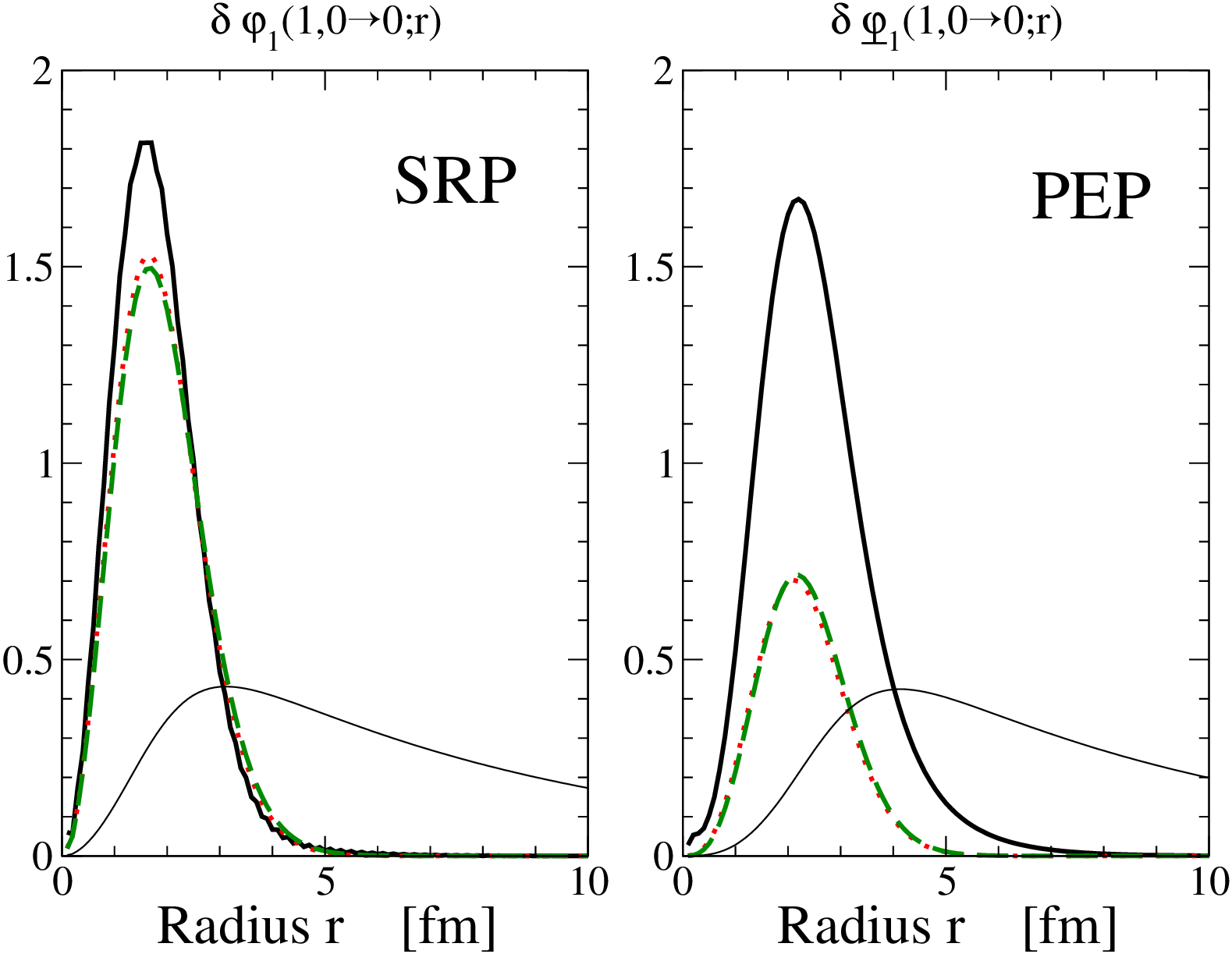}
\caption{(color online) The doorway state in the $l=0$ channel 
$\delta \varphi_1(0\rightarrow 0;r)$
produced by a Gaussian excitation for SRP and PEP transformations. The solid
line stands for the consistent excitation operator, dotted line for the
internal approximation and dashed line for the diagonal approximation. The
thin line stands for the initial wave function $\varphi _{1}^{0}(r)$.}
\label{figu13}
\end{figure}

In Fig.~\ref{figu13}, we represent $\delta \varphi_1(0\rightarrow 0;r)$ for SRP and PEP
transformations. The solid line stands for the consistent transformation of
the Gaussian interaction, the dotted line for its \textit{internal}
approximation and the dashed line for the \textit{diagonal}
approximation. The \textit{internal} approximation and the \textit{diagonal}
approximation give about the same results but are both very different from
the consistent calculation. These two figures illustrate the importance of
the off-diagonal terms which induce very different doorway states.

Summarizing our results, we can assert that
the Gaussian interaction is considerably modified by the SUSY
transformations and the \textit{internal} approximation is certainly a bad
approximation as it is illustrated in Fig.~\ref{figu13}. Hence, calculations
of structure properties and reaction mechanism which involve SUSY
transformations should never neglect the transformation of the excitation
operator (or residual interaction) for radii inside the core
potential.

\section{Conclusion}

In this article, we discussed a consistent framework to perform quantum 
mechanics SUSY
transformation to take into account the internal degrees of freedom in a
core approximation. This method is totally equivalent to the full
projector-method and is formally very interesting since it provides
justifications for effective core-nucleon interactions and nucleon-nucleon
residual interaction. 
In  this study, we have considered several
kinds of external fields and performed a consistent SUSY transformation.
The consistent SUSY transformation provides equivalent effective interactions
between composite particle systems and thus can be safely used to describe
nuclear structure and reaction of nuclei. The consistent transformation of
additional fields as well as the transformation of the wave functions
(or the observables) is usually neglected in the literature (internal 
approximation)and we have shown
that it is not always justified.

Our conclusions are the following: in the case of electromagnetic induced
transitions, consistent SUSY transformation conserves all selection rules 
while 
the \textit{internal} approximation violates it. Performing different 
comparisons we have shown that the discrepancies might be large, 
affecting the node structure of the doorway states and changing the 
transition probabilities by sizeable factors. Even the sum rules can 
be affected by a large percentage, e.g. 33\% for the 
energy weighted sum rule of the dipole excitation. Hence, the use of the 
\textit{internal} approximation for external excitation operator might 
be dangerous and the results obtained should be carefully discussed. 
But the main
discrepancies between the consistent calculation and its \textit{internal}
approximation appear for external fields which have a strong overlap with
the core potential. For instance, it is the case of the Gaussian interaction
centered at small distance ($r_{0}<r_{c}$).

In all the cases, we have shown that due to the SUSY transformation, the
off-diagonal terms of the external fields are often more important than the
diagonal ones. This forbids an approximation which would be to take into
account only the SUSY modification of the diagonal term. Hence, the SUSY
transformation has to be fully implemented in order to preserve the symmetry
of the original Hamiltonian. 

All the discussion related to the excitation operator is valid for 
the observables. Since the wave functions are transformed either they 
should be transformed back before evaluating average values or the 
observables should be also transformed before being applied on a 
transformed state.  

In conclusion, in this article, we have stressed the importance to keep the
consistency of the QM-SUSY framework if there is an overlap between the core
potential and the additional interactions (excitation operator or 
observables). For instance, in a recent article
Hesse et al.~\cite{Hes99} have performed the \textsl{internal} SUSY
approximation and they have shown that in order to reproduce the known
binding energies and radii of $^{6}$He, $^{11}$Li and $^{11}$Be halo nuclei,
a readjustment of the core-neutron interaction is required. This effect might
be induced by the \textsl{internal} SUSY approximation which treats
improperly the $r^{2}$ observable for the halo neutrons and 
the residual interaction between them. The
consistent SUSY framework would be a way to extract informations concerning
the neutron-neutron interaction in the halo because there is a unique
mapping between the original known interaction and the effective 
one which includes
consistent removal of core orbits.

\begin{acknowledgments}
We are grateful to Daniel Baye, Armen Sedrakian and Piet Van Isacker for 
helpful comments in the first version of this paper.
\end{acknowledgments}

\end{document}